\documentclass[a4paper,11pt]{article}
\pdfoutput=1 

\usepackage{jheppub} 

\usepackage[T1]{fontenc} 
\usepackage[utf8]{inputenc}

\usepackage{isomath}

\usepackage{physics}

\newcommand{\lap}{\laplacian{}}

\newcommand{\ii}{\mathrm{i}}
\newcommand{\TT}{\mathrm{T}}
\newcommand{\ee}{\mathrm{e}}

\usepackage{bbold}

\usepackage{tikz-feynman}
\tikzfeynmanset{compat=1.1.0}

\title{\boldmath Developing local RG: quantum RG and BFSS}

\author[a]{Jo\~ao F. Melo}
\author[a,b]{and Jorge E. Santos}

\affiliation[a]{Department of Applied Mathematics and Theoretical Physics, University of Cambridge, Wilberforce Road, Cambridge, CB3 0WA, UK}
\affiliation[b]{Institute for Advanced Study, Princeton, NJ 08540, USA}

\emailAdd{jfm54@cam.ac.uk}
\emailAdd{jss55@cam.ac.uk}

\abstract{In this paper we study various forms of RG and apply these to the BFSS model of $N$ coincident D0-branes. Firstly, as a warm-up, we perform standard Wilsonian RG, investigating the conditions under which supersymmetry is preserved along the flow. Next, we develop a local RG scheme such that the cutoff is spacetime dependent, which could have further applications to studying QFT in curved spacetime. Finally, we test the conjecture put forward in \cite{2014:Lee} that the method of quantum RG could be the mechanism responsible for the gauge/gravity duality by applying it to the BFSS model, which has a known gravitational dual. Although not entirely conclusive some questions are raised about the applicability of quantum RG as a description of the AdS/CFT correspondence.}

\begin{document} 
\maketitle
\flushbottom
\section{Introduction}
In its most precise form, the AdS/CFT correspondence is an equality of partition functions, where sources in the field theory side correspond to boundary conditions on the dynamical fields of the gravity side \cite{1999:Maldacena, 1998:Gubser, 1998:Witten, 2000:Aharony}. In the large $N$ limit on the field theory side, and in the classical limit on the gravity side, we get, roughly,

\begin{equation}
    \expval{\exp(\int\dd[d]{x}\mathcal{O}\phi^{(0)})}_\text{QFT}=\eval{\ee^{-S_\text{SuGra}}}_{\lim_{z\to0}\bar{\phi}(z,x)z^{\Delta-d}=\phi^{(0)}(x)}\,.
\end{equation}

We can then use this to calculate correlation functions on both sides. However, this is not the whole story. If we try to evaluate the classical action as it stands, with boundary conditions precisely on the boundary of AdS, we would get infinity. As is standard in QFT calculations, the way to deal with this infinity is to do renormalisation, \emph{i.e.} introducing counterterms to absorb the infinities. This procedure has been extensively developed, and is now a very standard technique under the name of Holographic Renormalisation \cite{1999:Balasubramanian,2001:deHaro,2002:Skenderis, 2008:Kanitscheider,Papadimitriou:2016yit}.

There are many interesting peculiarities with this idea. Firstly, it seems that what would normally be the UV divergences in standard QFT are in fact IR divergences in the gravity side. Further, what plays the role of renormalisation scale is in fact the radial direction in AdS spacetime. This is but one of the many hints that there is some deep connection between scale in the field theory side, and the radial direction in the gravity side \cite{1998:Akhmedov, 1999:Balasubramanian}.

Nonetheless, despite it's success, this also leaves many questions unanswered. The most immediate one is diffeomorphism invariance. What do we mean by radial direction? That is surely not a gauge invariant statement. Secondly, it's now very well known that renormalisation in QFT is not about removing annoying infinities, it's about coarse graining, integrating out degrees of freedom we do not have access to, in order to get a description relevant at our desired scale \cite{1974:Wilson}. Is there any way we can understand Holographic Renormalisation under a Wilsonian point of view?

As is to be expected, these questions have long been explored. It didn't take very long to understand that the would be RG flow in the gravity side is given by the Hamilton-Jacobi formulation, where instead of time evolution we consider radial evolution \cite{2000:Verlinde, 2000:Boer, 2001:Boer}. There have also been many proposals on how to give a more diffeomorphic invariant meaning to this radial direction \cite{2011:Heemskerk, 2011:Faulkner, 2012:Grozdanov, 2013:Balasubramanian, 2016:Lizana}. Most of which include interpreting different RG schemes on the QFT side as different coordinate systems in the bulk. What was in general poorly understood is which scheme corresponds to which coordinate system. More recently there is a proposal for the generic correspondence between smooth schemes on both sides \cite{2019:Ardalan}, and another for the particular case of dimensional regularisation \cite{2019:Bzowski}. As is to be expected, the relation between the two is not at all simple.

The difficulty with all these ideas (and the major interest) lies in the fact that, in order to get a full understanding of this issue, we would need to perform some sort of RG on the field theory side, and then compare with some sort of radial evolution on the gravity side, which, essentially, requires proving the conjecture. Conversely, we could also go the other way around, instead of thinking that it's a shame we need to prove the conjecture in order to answer these questions, we can try to answer these questions as a means to try to prove the conjecture. The goal of this paper is to do precisely that, to test, in a simple case, whether one of these proposals holds or not.

The proposal to be analysed in particular is the Quantum Renormalisation Group (QRG) \cite{2010:Lee, 2012:Lee, 2014:Lee}, which, briefly, consists of applying the following procedure to a QFT with matrix valued fields (this will be covered in more detail in section \ref{sec:overviewqrg}):

\begin{enumerate}
    \item Turn on single trace operator deformations with sources
    \item Do an infinitesimal local RG transformation
    \item Add auxiliary dynamical fields to project onto the space of single trace operators
    \item Iterate
\end{enumerate}

In this way, from a $d$-dim QFT we generate a $(d+1)$-dim action where what were sources are now dynamical fields. The proposal in \cite{2014:Lee} is that the new action would be the holographic dual to that CFT giving a concrete realisation of the AdS/CFT correspondence.\footnote{Other similar proposals include \cite{Papadimitriou:2004ap,Papadimitriou:2007sj,Papadimitriou:2010as,Papadimitriou:2011qb,2016:Behr1, 2016:Behr2,Mandal:2016rmt, 2016:Mukhopadhyay, 2017:Sathiapalan, 2019:Sathiapalan}, however, this paper will restrict its attention to testing QRG.}

Since the original paper, some follow-up work has been done, namely some hints for it's application to the original AdS$_5$/CFT$_4$ case \cite{2015:Nakayama}, a concrete calculation for the $U(N)$ vector model \cite{2015:Lunts}, and, understanding the conditions under which one can recover full $(d+1)$-dim diffeomorphism invariance \cite{2017:Shyam}. However, there has been no explicit calculation, starting from a QFT with a known gravitation dual, performing QRG and checking whether we end up with the same theory.

This is exactly what has been accomplished with this paper. The QFT chosen was the $\mathcal{N}=16$, one-dimensional super Yang-Mills theory with gauge group $SU(N)$, more commonly known as the BFSS model after the authors of \cite{1997:Banks}. This theory not only has a known gravitational dual \cite{1998:Itzhaki, 2000:Sekino, 2013:Ortiz, 2013:Anabalon}, but also is extremely simple given that it is one-dimensional, a fact which allows us to perform all calculations explicitly. In the end after we perform QRG the results seem to differ from the gravity predictions \cite{2000:Sekino,2014:Ortiz} (which have matched by lattice simulations \cite{2011:Hanada}). Even though QRG cannot be completely ruled out some questions are raised as to what would be needed to make it work or prove it wrong.


We begin section \ref{sec:rgbfss} by performing standard (\emph{i.e.} not quantum) RG on the BFSS model. This result by itself, as far as the authors are aware, is absent from the literature, mainly because there are no UV divergences, therefore, by itself, this is not very useful. However, it turns out it is a very useful playground to explore how one can break or preserve supersymmetry under an RG flow since we can compute everything explicitly. In section \ref{sec:lrg}, we address the first main concern, how to define a local version of RG. It turns out one can define this under certain restrictions, and we give a concrete example of how to achieve this. We have developed this formalism to apply to QRG, however, it may be interesting in it's own right, \emph{e.g.} if one wanted to perform RG in a curved background spacetime. Finally, in section \ref{sec:qrg}, we put everything together and perform QRG on the BFSS model. We start be reviewing the QRG procedure in detail and the holographic duality in BFSS. Then we go to the main calculations, highlighting the disagreement with known results.
\section{Renormalisation Group flow of BFSS model} \label{sec:rgbfss}
In this section we calculate the renormalisation group flow of the BFSS model in the case where the renormalisation scale is spacetime independent. We start by a brief review of the BFSS model, then we move on to the calculation using a hard momentum cutoff. Already here we find interesting ways to avoid breaking supersymmetry. After this prelude we discuss how to implement RG with a smooth cutoff in the sense of exact RG, we find that we always break supersymmetry in that case. Finally we give some remarks on (failed) attempts to circumvent the aforementioned supersymmetry breaking.
\subsection{Overview of the model}\label{subsec:reviewbfss}
The BFSS model is the maximally supersymmetry matrix quantum mechanics describing the dynamics of $N$ D0-branes. Equivalently, it is the $\mathcal{N}=16$ super Yang-Mills theory in $d=1$ dimensions with gauge group $SU(N)$, which can be obtain by dimensional reduction of the $\mathcal{N}=1$ super Yang-Mills in $d=10$ dimensions. It was originally introduced in \cite{1997:Banks} as a description of M-theory in the infinite momentum frame in the uncompactified limit, only later was its role in the gauge/gravity duality fully appreciated \cite{1998:Itzhaki}. For a general review of this model see \cite{2017:Ydri}.

This theory has an $SU(N)$ gauge field $A$, nine scalars $X_i$ ($i=1,\dots,9$), and 16 fermions $\psi_\alpha$ ($\alpha=1,\dots,16$). Both the scalars and the fermions are in the adjoint representation of the gauge group and therefore are represented by Hermitian, traceless, $N\times N$ matrices. The action for this model is (in Euclidean time):

\begin{align}
    S[A,X,\psi]=\frac{N}{\lambda}\int\dd{\tau}\Tr\bigg\{&\frac{1}{2}(\mathrm{D}_\tau X_i)^2+\frac{1}{2}\psi_\alpha \mathrm{D}_\tau \psi_\alpha+\nonumber\\
    &+\frac{1}{2}\psi_\alpha(\gamma_i)_{\alpha\beta}[X_i, \psi_\beta]-\frac{1}{4}[X_i,X_j]^2\bigg\}
    \label{eq:bfssaction}
\end{align}
where $\lambda=Ng_\text{YM}^2$ is the usual 't Hooft coupling. We are using the convention where the generators of the Lie algebra are Hermitian, and therefore they obey
\begin{equation}
[T^a,T^b]=\ii f_{abc}T^c\,.
\end{equation}
Furthermore, we normalise $T$ as $\Tr(T^aT^b)=\delta^{ab}$. The covariant derivative in Eq.~(\ref{eq:bfssaction}) acts as $\mathrm{D}_\tau=\partial_\tau+\ii[A,\cdot]$. Finally, $\gamma_i$ are the nine-dimensional Dirac gamma matrices, which are real, symmetric matrices satisfying $\{\gamma_i,\gamma_j\}=2\delta_{ij}$.

As mentioned above, this theory is invariant under a supersymmetry transformation with 16 supercharges whose precise form will not be relevant for the subsequent discussion. Note also that the gauge field is not dynamical, therefore we can completely fix the gauge with $A=0$ without the need to introduce Fadeev-Popov ghosts. This is one of the many simplifying aspects of the theory. In the remainder of the manuscript we assume we are in such a gauge.

It will also prove to be useful to do the rescaling,

\begin{equation}
    \tilde{X}_i=\sqrt{\frac{N}{\lambda}}X_i,~~~~~\tilde{\psi}_\alpha=\sqrt{\frac{N}{\lambda}}\psi_\alpha
    \label{eq:rescale}
\end{equation}
so that, in these new variables, the action looks like,

\begin{align}
    S[X,\psi]=\int\dd{\tau}\Tr\bigg\{&\frac{1}{2}(\partial_\tau \tilde{X}_i)^2+\frac{1}{2}\tilde{\psi}_\alpha \partial_\tau \tilde{\psi}_\alpha+\nonumber\\
    &+\frac{1}{2}\sqrt{\frac{\lambda}{N}}\tilde{\psi}_\alpha(\gamma_i)_{\alpha\beta}[\tilde{X}_i, \tilde{\psi}_\beta]-\frac{\lambda}{4N}[\tilde{X}_i,\tilde{X}_j]^2\bigg\}\,.
\end{align}

We note that in the large $N$ limit $N\to\infty$, the original untilded variables are $\mathcal{O}(N^0)$. Finally, in order to do the perturbative calculations presented in the subsequent sections, it is convenient to write the action in terms of the structure constants,

\begin{align}
    S[X,\psi]=\int\dd{\tau}\bigg\{&\frac{1}{2}(\partial_\tau \tilde{X}_i^a)^2+\frac{1}{2}\tilde{\psi}_\alpha^a \partial_\tau \tilde{\psi}_\alpha^a+\ii\frac{1}{2}\sqrt{\frac{\lambda}{N}}(\gamma_i)_{\alpha\beta} f_{abc}\tilde{\psi}_\alpha^a\tilde{X}_i^b \tilde{\psi}_\beta^c+\nonumber\\
    &+\frac{\lambda}{4N}f_{abe}f_{cde}\tilde{X}_i^a\tilde{X}_j^b\tilde{X}_i^c\tilde{X}_j^d\bigg\}\,.
    \label{eq:structure}
\end{align}

From Eq.(\ref{eq:structure}) we can easily read the associated Feynman rules:

\begin{itemize}
    \item Scalar propagator:
    \begin{equation*}
    \begin{tikzpicture}[baseline=(a)]
      \begin{feynman}
        \vertex (a) {$i,a$};
        \vertex [right=2.5cm of a] (b) {$j,b$};
        
        \diagram* {
          (a) -- [scalar,momentum={$p$}] (b),
        };
      \end{feynman}
    \end{tikzpicture}
    =\frac{\delta_{ij}\delta_{ab}}{p^2}
    \end{equation*}
    \item Fermion propagator:
    \begin{equation*}
    \begin{tikzpicture}[baseline=(a)]
      \begin{feynman}
        \vertex (a) {$\alpha,a$};
        \vertex [right=2.5cm of a] (b) {$\beta,b$};
        
        \diagram* {
          (a) -- [plain,momentum={$p$}] (b),
        };
      \end{feynman}
    \end{tikzpicture}
    =\frac{\delta_{\alpha\beta}\delta_{ab}}{p}
    \end{equation*}    
    \item Cubic coupling:
    \begin{equation*}
    \begin{tikzpicture}[baseline=(centre)]
      \begin{feynman}
        \vertex (a) {$\alpha,a$};
        \vertex [right=1.5cm of a] (centre);
        \vertex [right=1cm of centre] (aux);
        \vertex [below=1cm of aux] (b) {$\beta,b$};
        \vertex [above=1cm of aux] (c) {$i,c$};
        
        \diagram* {
          (a) -- [plain] (centre),
          (centre) -- [plain] (b),
          (centre) -- [scalar] (c),
        };
      \end{feynman}
    \draw[fill=black] (centre) circle (1.5pt);
    \end{tikzpicture}
    =-\ii\sqrt{\frac{\lambda}{N}}(\gamma_i)_{\alpha\beta}f_{cba}
    \end{equation*} 
    \item Quartic coupling:
    \begin{equation*}
    \begin{tikzpicture}[baseline=(centre)]
      \begin{feynman}
        \vertex (a) {$i,a$};
        \vertex [right=2.5cm of a] (b) {$j,b$};
        \vertex [below=2.5cm of a] (c) {$k,c$};
        \vertex [below=2.5cm of b] (d) {$l,d$};
        \vertex [below=1.25cm of a] (aux1);
        \vertex [right=1.25cm of aux1] (centre);
        
        \diagram* {
          (a) -- [scalar] (centre),
          (b) -- [scalar] (centre),
          (c) -- [scalar] (centre),
          (d) -- [scalar] (centre),
        };
      \end{feynman}
    \draw[fill=black] (centre) circle (1.5pt);
    \end{tikzpicture}
    \begin{array}{rl}
    =-\displaystyle\frac{\lambda}{N}\big[ & f_{abe}f_{cde}(\delta_{ik}\delta_{jl}-\delta_{il}\delta_{jk})+\\
    + & f_{ace}f_{bde}(\delta_{ij}\delta_{kl}-\delta_{il}\delta_{jk})+\\
    + & f_{ade}f_{bce}(\delta_{ij}\delta_{kl}-\delta_{ik}\delta_{jl})\big]
    \end{array}
    \end{equation*}     
\end{itemize}

\subsection{RG with a hard momentum cutoff}\label{subsec:hardrg}

As a warm-up calculation, we start by computing the perturbative 1-loop RG flow of this model. Since this is a one-dimensional theory, there will be an infinite number of relevant interactions that will be turned on by the RG flow, rendering our perturbative approximation useless. We will, nonetheless, proceed with the calculations and only consider diagrams with up to four external legs. This is completely artificial and unjustified, however, we will proceed with this calculation because there are still some interesting lessons to take from this analysis to do with supersymmetry.

We will impose a hard momentum cutoff by demanding that our fields only have support for momenta $\abs{p}<\Lambda_0$. Then, to lower the cutoff, we integrate over modes with support in momentum space $\Lambda<\abs{p}<\Lambda_0$. The calculations themselves involve rather tedious index manipulations, for that reason we relegate the details to appendix \ref{app:oneloop} and only present the main results in the core text. The relevant diagrams at 1-loop order and up to four external fields are (where we denote the high energy modes with blue):

\paragraph{Tadpole}

This one is trivially zero by the index structure.

\begin{equation*}
\begin{tikzpicture}[baseline=(a)]
  \begin{feynman}
    \vertex (a);
    \vertex [right=1.5cm of a] (centre);
    \vertex [right=0.5cm of centre] (aux1);
    \vertex [above=0.5cm of aux1] (aux2);
    \vertex [right=0.5cm of aux1] (aux3);
    \vertex [below=0.5cm of aux1] (aux4);
    
    \diagram* {
      (a) -- [scalar] (centre),
      (centre) -- [blue,plain,out=90,in=180] (aux2),
      (aux2) -- [blue,plain,out=0,in=90] (aux3),
      (aux3) -- [blue,plain,out=-90,in=0] (aux4),
      (aux4) -- [blue,plain,out=180,in=-90] (centre),
    };
  \end{feynman}
  \draw[fill=black] (centre) circle (1.5pt);
\end{tikzpicture}
=0\,.
\end{equation*}

\paragraph{Scalar propagator} There are two diagrams that contribute. We can either have scalar loop

\begin{equation*}
\begin{tikzpicture}[baseline=(b.south)]
  \begin{feynman}
    \vertex (a) {$i,a$};
    \vertex [right=1.5cm of a] (centre);
    \vertex [right=1.25cm of centre] (b) {$j,b$};
    \vertex [above=1.5cm of centre] (aux);
    
    \diagram* {
      (a) -- [scalar,out=0,in=180,momentum'={[arrow shorten=0.25] $p$}] (centre),
      (centre) -- [scalar,out=0,in=180,momentum'={[arrow shorten=0.25] $p$}] (b),
      (centre) -- [blue,scalar,out=135,in=180,momentum={[arrow style=blue, arrow shorten=0.27] $\omega$}] (aux),
      (aux) -- [blue,scalar,out=0,in=45] (centre),
    };
    \draw[fill=black] (centre) circle (1.5pt);
  \end{feynman}
\end{tikzpicture}
=-16\lambda\delta_{ab}\delta_{ij}\int_{\abs{\omega}\in [\Lambda,\Lambda_0]}\frac{\dd{\omega}}{2\pi}\frac{1}{\omega^2}=-\frac{16\lambda}{\pi}\delta_{ij}\delta_{ab}\qty(\frac{1}{\Lambda}-\frac{1}{\Lambda_0})\,,
\end{equation*}
or a fermionic loop
\begin{equation}
\begin{tikzpicture}[baseline=(a)]
  \begin{feynman}
    \vertex (a) {$i,a$};
    \vertex [right=1.4cm of a] (centre1);
    \vertex [right=0.6cm of centre1] (aux1);
    \vertex [above=0.6cm of aux1] (aux2);
    \vertex [below=0.6cm of aux1] (aux3);
    \vertex [right=0.6cm of aux1] (centre2);
    \vertex [right=1.1cm of centre2] (b) {$j,b$};
    
    \diagram* {
      (a) -- [scalar,momentum'={[arrow shorten=0.25]$p$}] (centre1),
      (centre1) -- [blue,plain,out=90,in=180,momentum={[arrow style=blue,arrow shorten=0.27]$\omega$}] (aux2),
      (aux2) -- [blue,plain,out=0,in=90] (centre2),
      (centre2) -- [blue,plain,out=-90,in=0,momentum={[arrow style=blue,arrow shorten=0.27]$\omega-p$}] (aux3),
      (aux3) -- [blue,plain,out=180,in=-90] (centre1),
      (centre2) -- [scalar,momentum={[arrow shorten=0.25]$p$}] (b),
    };
  \end{feynman}
  \draw[fill=black] (centre1) circle (1.5pt);
  \draw[fill=black] (centre2) circle (1.5pt);
\end{tikzpicture}
=16\lambda\delta_{ab}\delta_{ij}\int\frac{\dd{\omega}}{2\pi}\frac{1}{\omega(\omega-p)}\,.
\label{eq:scalarprop1}
\end{equation}

For the scalar mode we must have $\abs{\omega}\in[\Lambda,\Lambda_0]$. For the fermionic mode, one might naively think that the region of integration is also $\abs{\omega}\in[\Lambda,\Lambda_0]$, just as for the scalar. However, that would be wrong. In fact there is also a high energy mode with momentum $\omega-p$ so, since that mode only has support when \textit{its} momentum is in the range $[\Lambda,\Lambda_0]$ we must also impose that $\abs{\omega-p}\in[\Lambda,\Lambda_0]$. Usually, integrating over these intricate regions is prohibitively difficult, however, for one-dimensional integrals, they can be done analytically. If we do not integrate over this region, we get non-sensical answers. For instance, the answer would depend on which line of the loop we give momentum $\omega$ and which line we give momentum $\omega-p$\footnote{For this diagram that does not happen because the two lines are identical, but further ahead one we would see such an effect.}.

Let us define
\begin{equation}
I\equiv\{\omega|~\abs{\omega}\in[\Lambda,\Lambda_0]\land\abs{\omega-p}\in[\Lambda,\Lambda_0]\}
\label{eq:In}
\end{equation}
which brings Eq.~(\ref{eq:scalarprop1}) to
\begin{equation*}
    16\lambda\delta_{ab}\delta_{ij}\int_I\frac{\dd{\omega}}{2\pi}\frac{1}{\omega(\omega-p)}=\begin{cases}
    \displaystyle\frac{16\lambda\delta_{ij}\delta_{ab}}{p\pi}\log\left[\frac{(\Lambda+p)(\Lambda_0-p)}{\Lambda\Lambda_0}\right],~~~p>0
    \\
    \\
    \displaystyle\frac{16\lambda\delta_{ij}\delta_{ab}}{p\pi}\log\left[\frac{\Lambda\Lambda_0}{(\Lambda-p)(\Lambda_0+p)}\right],~~~p<0
    \end{cases}\,.
\end{equation*}
Expanding in powers of $p$ yields
\begin{equation*}
    \frac{16\lambda\delta_{ij}\delta_{ab}}{\pi}\qty[\qty(\frac{1}{\Lambda}-\frac{1}{\Lambda_0})-\frac{\abs{p}}{2}\qty(\frac{1}{\Lambda_0^2}+\frac{1}{\Lambda^2})+\frac{p^2}{3}\qty(\frac{1}{\Lambda^3}-\frac{1}{\Lambda_0^3})+\mathcal{O}(p^3)]\,.
\end{equation*}
There is a linear term in $p$ which could be worrisome, however, in $d=1$ this is a total derivative, so we shall drop it. Note that the would-be mass term cancels between the two diagrams and we are left with just a wavefunction renormalisation contribution.

\begin{equation}
    Z'_X=1-\frac{16\lambda}{3\pi}\qty(\frac{1}{\Lambda^3}-\frac{1}{\Lambda_0^3})\,.
\end{equation}

\paragraph{Fermion propagator} There is only one diagram that contributes:

\begin{equation*}
\begin{tikzpicture}[baseline=(a)]
  \begin{feynman}
    \vertex (a) {$\alpha,a$};
    \vertex [right=1.5cm of a] (centre1);
    \vertex [right=1.5cm of centre1] (centre2);
    \vertex [right=1.1cm of centre2] (b) {$\beta,b$};
    \vertex [right=0.75cm of centre1] (aux1);
    \vertex [above=0.75cm of aux1] (aux2);
    
    \diagram* {
      (a) -- [plain,momentum'={[arrow shorten=0.25]$p$}] (centre1),
      (centre1) -- [blue,plain,momentum'={[arrow style=blue,arrow shorten=0.25]$p-\omega$}] (centre2),
      (centre2) -- [plain,momentum'={[arrow shorten=0.25]$p$}] (b),
      (centre1) -- [blue,scalar,out=90,in=180,momentum={[arrow style=blue,arrow shorten=0.25]$\omega$}] (aux2),
      (aux2) -- [blue,scalar,out=0,in=90] (centre2),
    };
  \end{feynman}
  \draw[fill=black] (centre1) circle (1.5pt);
  \draw[fill=black] (centre2) circle (1.5pt);
\end{tikzpicture}
=18\lambda\delta_{ab}\delta_{\alpha\beta}\int_I\frac{\dd{\omega}}{2\pi}\frac{1}{\omega^2(p-\omega)}
\end{equation*}
where once again we have to be careful about the integration region and integrate over $I$ as defined in Eq.~(\ref{eq:In}):
\begin{equation}
    \begin{cases}
        \displaystyle\frac{9\lambda\delta_{ab}\delta_{\alpha\beta}}{p^2\pi}\qty[p\qty(\frac{1}{\Lambda}-\frac{1}{\Lambda_0}+\frac{1}{\Lambda+p}-\frac{1}{\Lambda_0-p})+2\log(\frac{\Lambda\Lambda_0}{(\Lambda+p)(\Lambda_0-p)})],~p>0
        \\
        \\
        \displaystyle\frac{9\lambda\delta_{ab}\delta_{\alpha\beta}}{p^2\pi}\qty[p\qty(\frac{1}{\Lambda_0}-\frac{1}{\Lambda}+\frac{1}{\Lambda_0+p}-\frac{1}{\Lambda-p})+2\log(\frac{\Lambda\Lambda_0}{(\Lambda_0+p)(\Lambda-p)})],~p<0
    \end{cases}\,.
\end{equation}
Expanding in powers of $p$ we get,
\begin{equation}
 \frac{3\lambda\delta_{ab}\delta_{\alpha\beta}}{\pi}\abs{p}\qty(\frac{1}{\Lambda^3}-\frac{1}{\Lambda_0^3})\,,
\end{equation}
which gives a wavefunction renormalisation of
\begin{equation}
    Z'_\psi=1-\frac{3\lambda}{\pi}\qty(\frac{1}{\Lambda^3}-\frac{1}{\Lambda_0^3})\,.
\end{equation}

\paragraph{Triangle diagram} This is also trivially zero by the index structure

\begin{equation*}
\begin{tikzpicture}[baseline=(a)]
  \begin{feynman}
    \vertex (a);
    \vertex [right=1.25cm of a] (centre1);
    \vertex [right=1cm of centre1] (aux1);
    \vertex [above=0.75cm of aux1] (centre2);
    \vertex [below=0.75cm of aux1] (centre3);
    \vertex [right=1.25cm of centre2] (b);
    \vertex [right=1.25cm of centre3] (c);
    
    \diagram* {
      (a) -- [scalar] (centre1),
      (centre1) -- [blue,plain] (centre2),
      (centre2) -- [blue,plain] (centre3),
      (centre1) -- [blue,plain] (centre3),
      (centre2) -- [scalar] (b),
      (centre3) -- [scalar] (c),
    };
  \end{feynman}
  \draw[fill=black] (centre1) circle (1.5pt);
  \draw[fill=black] (centre2) circle (1.5pt);
  \draw[fill=black] (centre3) circle (1.5pt);
\end{tikzpicture}
=0\,.
\end{equation*}

\paragraph{Cubic coupling} There is only one diagram that contributes

\begin{equation*}
\begin{tikzpicture}[baseline=(c)]
  \begin{feynman}
    \vertex (c) {$i,c$};
    \vertex [right=1.75cm of c] (centre1);
    \vertex [right=0.9cm of centre1] (aux1);
    \vertex [right=1.1cm of aux1] (aux2);
    \vertex [above=0.65cm of aux1] (centre2);
    \vertex [below=0.675cm of aux1] (centre3);
    \vertex [above=1.2cm of aux2] (a) {$\alpha,a$};
    \vertex [below=1.2cm of aux2] (b) {$\beta,b$};

    \diagram* {
      (c) -- [scalar,momentum={[arrow shorten=0.25]$p_3$}] (centre1);
      (centre1) -- [blue,plain,rmomentum={[arrow style=blue,arrow shorten=0.25]$\omega$}] (centre2);
      (centre2) -- [plain,rmomentum={[arrow shorten=0.25]$p_1$}] (a);
      (centre1) -- [blue,plain,momentum'={[arrow style=blue,arrow shorten=0.25]$\omega+p_3$}] (centre3);
      (centre3) -- [plain,rmomentum'={[arrow shorten=0.25]$p_2$}] (b);
      (centre2) -- [blue,scalar,rmomentum={[arrow style=blue,arrow shorten=0.25]$\omega-p_1$}] (centre3);
    };
  \end{feynman}
  \draw[fill=black] (centre1) circle (1.5pt);
  \draw[fill=black] (centre2) circle (1.5pt);
  \draw[fill=black] (centre3) circle (1.5pt);
\end{tikzpicture}
=7\ii\lambda\sqrt{\frac{\lambda}{N}}(\gamma_i)_{\alpha\beta}f_{acb}\int\frac{\dd{\omega}}{2\pi}\frac{1}{\omega(\omega+p_3)(\omega-p_1)^2}\,.
\end{equation*}
Since we just want the correction to the cubic coupling we will set the external momenta to zero. This also means there are no subtleties with the region of integration. We then get for the correction to the cubic coupling,
\begin{equation}
    \lambda'_{(3)}=\sqrt{\frac{\lambda}{N}}\qty[1-7\lambda\int\frac{\dd{\omega}}{2\pi}\frac{1}{\omega^4}]=\sqrt{\frac{\lambda}{N}}\qty[1-\frac{7}{3\pi}\lambda\qty(\frac{1}{\Lambda^3}-\frac{1}{\Lambda_0^3})]\,.
\end{equation}

\paragraph{Quartic coupling} Now there are six diagrams that contribute at 1-loop order, they are all distinct and rather messy. However, setting the external momenta to zero allows us to add up all these diagrams to get something nice in the end. After the dust settles the correction to the quartic couplic is:

\begin{equation*}
\begin{tikzpicture}[baseline=(centre1)]
  \begin{feynman}
    \vertex (a);
    \vertex [below=1.5cm of a] (c);
    \vertex [below=0.75cm of a] (aux1);
    \vertex [right=2.5cm of a] (b);
    \vertex [right=2.5cm of c] (d);
    \vertex [right=0.75cm of aux1] (centre1);
    \vertex [right=1cm of centre1] (centre2);
    \vertex [right=0.5cm of centre1] (aux2);
    \vertex [above=0.5cm of aux2] (aux3);
    \vertex [below=0.5cm of aux2] (aux4);
    
    \diagram* {
      (a) -- [scalar,out=-45,in=135] (centre1),
      (c) -- [scalar,out=45,in=-135] (centre1),
      (centre1) -- [blue,scalar,out=90,in=180] (aux3),
      (centre1) -- [blue,scalar,out=-90,in=180] (aux4),
      (aux3) -- [blue,scalar,out=0,in=90] (centre2),
      (aux4) -- [blue,scalar,out=0,in=-90] (centre2),
      (centre2) -- [scalar,out=45,in=-135] (b),
      (centre2) -- [scalar,out=-45,in=135] (d),
    };
  \end{feynman}
  \draw[fill=black] (centre1) circle (1.5pt);
  \draw[fill=black] (centre2) circle (1.5pt);
\end{tikzpicture}
+
\begin{tikzpicture}[baseline=(aux2)]
  \begin{feynman}
    \vertex (a);
    \vertex [below=2.5cm of a] (c);
    \vertex [right=1.5cm of a] (b);
    \vertex [right=1.5cm of c] (d);
    \vertex [right=0.75cm of a] (aux1);
    \vertex [below=0.75cm of aux1] (centre1);
    \vertex [below=1cm of centre1] (centre2);
    \vertex [below=0.5cm of centre1] (aux2);
    \vertex [left=0.5cm of aux2] (aux3);
    \vertex [right=0.5cm of aux2] (aux4);
    
    \diagram* {
      (a) -- [scalar,out=-45,in=135] (centre1),
      (c) -- [scalar,out=45,in=-135] (centre2),
      (centre1) -- [scalar,out=45,in=-135] (b),
      (centre2) -- [scalar,out=-45,in=135] (d),
      (centre1) -- [blue,scalar,out=180,in=90] (aux3),
      (aux3) -- [blue,scalar,out=-90,in=180] (centre2),
      (centre1) -- [blue,scalar,out=0,in=90] (aux4),
      (aux4) -- [blue,scalar,out=-90,in=0] (centre2),
    };
  \end{feynman}
  \draw[fill=black] (centre1) circle (1.5pt);
  \draw[fill=black] (centre2) circle (1.5pt);
\end{tikzpicture}
+
\begin{tikzpicture}[baseline=(aux2)]
  \begin{feynman}
    \vertex (a);
    \vertex [below=2.5cm of a] (c);
    \vertex [right=2cm of a] (b);
    \vertex [right=2cm of c] (d);
    \vertex [right=0.75cm of a] (aux1);
    \vertex [below=0.75cm of aux1] (centre1);
    \vertex [below=1cm of centre1] (centre2);
    \vertex [below=0.5cm of centre1] (aux2);
    \vertex [left=0.5cm of aux2] (aux3);
    \vertex [right=0.5cm of aux2] (aux4);
    
    \diagram* {
      (a) -- [scalar,out=-45,in=135] (centre1),
      (c) -- [scalar,out=45,in=-135] (centre2),
      (centre1) -- [scalar,out=5,in=135] (d),
      (centre2) -- [scalar,out=-5,in=-135] (b),
      (centre1) -- [blue,scalar,out=180,in=90] (aux3),
      (aux3) -- [blue,scalar,out=-90,in=180] (centre2),
      (centre1) -- [blue,scalar,out=0,in=90] (aux4),
      (aux4) -- [blue,scalar,out=-90,in=0] (centre2),
    };
  \end{feynman}
  \draw[fill=black] (centre1) circle (1.5pt);
  \draw[fill=black] (centre2) circle (1.5pt);
\end{tikzpicture}
+
\end{equation*}

\begin{equation*}
+    
\begin{tikzpicture}[baseline=(aux2)]
  \begin{feynman}
    \vertex (a);
    \vertex [right=2.5cm of a] (b);
    \vertex [below=2.5cm of a] (c);
    \vertex [right=2.5cm of c] (d);
    \vertex [right=1.25cm of a] (aux1);
    \vertex [below=1.25cm of aux1] (aux2);
    \vertex [above=0.5cm of aux2] (aux3);
    \vertex [left=0.5cm of aux3] (centre1);
    \vertex [right=0.5cm of aux3] (centre2);
    \vertex [below=0.5cm of aux2] (aux4);
    \vertex [left=0.5cm of aux4] (centre3);
    \vertex [right=0.5cm of aux4] (centre4);
    
    \diagram* {
        (a) -- [scalar] (centre1),
        (b) -- [scalar] (centre2),
        (c) -- [scalar] (centre3),
        (d) -- [scalar] (centre4),
        (centre1) -- [blue,plain] (centre2),
        (centre1) -- [blue,plain] (centre3),
        (centre2) -- [blue,plain] (centre4),
        (centre3) -- [blue,plain] (centre4),
    };
  \end{feynman}
  \draw[fill=black] (centre1) circle (1.5pt);
  \draw[fill=black] (centre2) circle (1.5pt);
  \draw[fill=black] (centre3) circle (1.5pt);
  \draw[fill=black] (centre4) circle (1.5pt);
\end{tikzpicture}
+
\begin{tikzpicture}[baseline=(aux2)]
  \begin{feynman}
    \vertex (a);
    \vertex [right=2.5cm of a] (b);
    \vertex [below=2.5cm of a] (c);
    \vertex [right=2.5cm of c] (d);
    \vertex [right=1.25cm of a] (aux1);
    \vertex [below=1.25cm of aux1] (aux2);
    \vertex [above=0.5cm of aux2] (aux3);
    \vertex [left=0.5cm of aux3] (centre1);
    \vertex [right=0.5cm of aux3] (centre2);
    \vertex [below=0.5cm of aux2] (aux4);
    \vertex [left=0.5cm of aux4] (centre3);
    \vertex [right=0.5cm of aux4] (centre4);
    
    \diagram* {
        (a) -- [scalar] (centre1),
        (b) -- [scalar] (centre4),
        (c) -- [scalar] (centre3),
        (d) -- [scalar] (centre2),
        (centre1) -- [blue,plain] (centre2),
        (centre1) -- [blue,plain] (centre3),
        (centre2) -- [blue,plain] (centre4),
        (centre3) -- [blue,plain] (centre4),
    };
  \end{feynman}
  \draw[fill=black] (centre1) circle (1.5pt);
  \draw[fill=black] (centre2) circle (1.5pt);
  \draw[fill=black] (centre3) circle (1.5pt);
  \draw[fill=black] (centre4) circle (1.5pt);
\end{tikzpicture}
+    
\begin{tikzpicture}[baseline=(aux2)]
  \begin{feynman}
    \vertex (a);
    \vertex [right=2.5cm of a] (b);
    \vertex [below=2.5cm of a] (c);
    \vertex [right=2.5cm of c] (d);
    \vertex [right=1.25cm of a] (aux1);
    \vertex [below=1.25cm of aux1] (aux2);
    \vertex [above=0.5cm of aux2] (aux3);
    \vertex [left=0.5cm of aux3] (centre1);
    \vertex [right=0.5cm of aux3] (centre2);
    \vertex [below=0.5cm of aux2] (aux4);
    \vertex [left=0.5cm of aux4] (centre3);
    \vertex [right=0.5cm of aux4] (centre4);
    
    \diagram* {
        (a) -- [scalar] (centre2),
        (b) -- [scalar] (centre1),
        (c) -- [scalar] (centre3),
        (d) -- [scalar] (centre4),
        (centre1) -- [blue,plain] (centre2),
        (centre1) -- [blue,plain] (centre3),
        (centre2) -- [blue,plain] (centre4),
        (centre3) -- [blue,plain] (centre4),
    };
  \end{feynman}
  \draw[fill=black] (centre1) circle (1.5pt);
  \draw[fill=black] (centre2) circle (1.5pt);
  \draw[fill=black] (centre3) circle (1.5pt);
  \draw[fill=black] (centre4) circle (1.5pt);
\end{tikzpicture}
\end{equation*}
\begin{equation}
    \lambda'_{(4)}=\frac{\lambda}{N}\qty[1+\frac{4\lambda}{3\pi}\qty(\frac{1}{\Lambda^3}-\frac{1}{\Lambda_0^3})]\,.
\end{equation}

Putting everything together, that is, including the wavefunction renormalisation and classical scaling into account we find, to leading order in $\lambda$,

\begin{align}
    \lambda_{(3)}(\Lambda)&=\sqrt{\frac{\Lambda_0}{\Lambda}}\sqrt{\frac{\lambda}{N}}\qty[1+\frac{10\lambda}{3\pi}\qty(\frac{1}{\Lambda^3}-\frac{1}{\Lambda_0^3})]\\
    \lambda_{(4)}(\Lambda)&=\frac{\Lambda_0}{\Lambda}\frac{\lambda}{N}\qty[1+\frac{12\lambda}{\pi}\qty(\frac{1}{\Lambda^3}-\frac{1}{\Lambda_0^3})]\,.
\end{align}

Even though we have not generated anything as egregious as a mass term for either the fermions or the scalars, the contribution to the cubic and quartic couplings is not quite right. At the quantum level, with this regulator, $\lambda_{(4)}\neq\lambda_{(3)}^2$ which signals a breaking of supersymmetry. 

By themselves, these results are not very surprising. In this theory, the supersymmetry algebra only closes on-shell, so a hard momentum cutoff will necessarily break supersymmetry (the next section will delve deeper into this issue). However, we have noticed a somewhat bizarre feature for which the interpretation is still not entirely clear (which is the main reason for including these calculations in the final manuscript). We can preserve supersymmetry at the 1-loop level if we prescribe the integration in a slightly different way. Instead of integrating with the physical constraint that all internal lines are high energy, we tried using the Feynman parameter method, which is usually used to combine propagators and make integrals more tractable (in our case we can do the calculation in both ways and compare the final answer). We then impose that the final integral is the one that sits in the range $[\Lambda,\Lambda_0]$. Like we previously mentioned, this is physically rather dubious, but it corresponds to the standard practice in higher dimensions (see for instance \cite{1995:Peskin}), and, surprisingly enough, it appears to preserve supersymmetry. 

The only diagrams that change are the contribution to the scalar propagator with a fermionic loop and the fermionic propagator. The contribution to the scalar propagator with a fermionic loop now yields
\begin{equation*}
\begin{tikzpicture}[baseline=(a)]
  \begin{feynman}
    \vertex (a) {$i,a$};
    \vertex [right=1.4cm of a] (centre1);
    \vertex [right=0.6cm of centre1] (aux1);
    \vertex [above=0.6cm of aux1] (aux2);
    \vertex [below=0.6cm of aux1] (aux3);
    \vertex [right=0.6cm of aux1] (centre2);
    \vertex [right=1.1cm of centre2] (b) {$j,b$};
    
    \diagram* {
      (a) -- [scalar,momentum'={[arrow shorten=0.25]$p$}] (centre1),
      (centre1) -- [blue,plain,out=90,in=180,momentum={[arrow style=blue,arrow shorten=0.27]$\omega$}] (aux2),
      (aux2) -- [blue,plain,out=0,in=90] (centre2),
      (centre2) -- [blue,plain,out=-90,in=0,momentum={[arrow style=blue,arrow shorten=0.27]$\omega-p$}] (aux3),
      (aux3) -- [blue,plain,out=180,in=-90] (centre1),
      (centre2) -- [scalar,momentum={[arrow shorten=0.25]$p$}] (b),
    };
  \end{feynman}
  \draw[fill=black] (centre1) circle (1.5pt);
  \draw[fill=black] (centre2) circle (1.5pt);
\end{tikzpicture}
=16\lambda\delta_{ab}\delta_{ij}\int\frac{\dd{\omega}}{2\pi}\frac{1}{\omega(\omega-p)}=
\end{equation*}
\begin{align}
    =&16\lambda\delta_{ab}\delta_{ij}\int_0^1\dd{x}\int\frac{\dd{\omega}}{2\pi}\frac{1}{(\omega-xp)^2}=\nonumber\\
    =&16\lambda\delta_{ab}\delta_{ij}\int_0^1\dd{x}\int_{\abs{l}\in[\Lambda,\Lambda_0]}\frac{\dd{l}}{2\pi}\frac{1}{l^2}=\nonumber\\
    =&\frac{16\lambda}{\pi}\delta_{ij}\delta_{ab}\qty(\frac{1}{\Lambda}-\frac{1}{\Lambda_0})\,,
\end{align}
which precisely cancels the contribution from the scalar loop, meaning there is no scalar wavefunction renormalisation with this regulator.

Finally the fermionic propagator becomes,
\begin{equation*}
\begin{tikzpicture}[baseline=(a)]
  \begin{feynman}
    \vertex (a) {$\alpha,a$};
    \vertex [right=1.5cm of a] (centre1);
    \vertex [right=1.5cm of centre1] (centre2);
    \vertex [right=1.1cm of centre2] (b) {$\beta,b$};
    \vertex [right=0.75cm of centre1] (aux1);
    \vertex [above=0.75cm of aux1] (aux2);
    
    \diagram* {
      (a) -- [plain,momentum'={[arrow shorten=0.25]$p$}] (centre1),
      (centre1) -- [blue,plain,momentum'={[arrow style=blue,arrow shorten=0.25]$p-\omega$}] (centre2),
      (centre2) -- [plain,momentum'={[arrow shorten=0.25]$p$}] (b),
      (centre1) -- [blue,scalar,out=90,in=180,momentum={[arrow style=blue,arrow shorten=0.25]$\omega$}] (aux2),
      (aux2) -- [blue,scalar,out=0,in=90] (centre2),
    };
  \end{feynman}
  \draw[fill=black] (centre1) circle (1.5pt);
  \draw[fill=black] (centre2) circle (1.5pt);
\end{tikzpicture}
=18\lambda\delta_{ab}\delta_{\alpha\beta}\int\frac{\dd{\omega}}{2\pi}\frac{1}{\omega^2(p-\omega)}=
\end{equation*}

\begin{align}
    =&18\lambda\delta_{ab}\delta_{\alpha\beta}\int_0^1\dd{x}\int\frac{\dd{\omega}}{2\pi}\frac{p-\omega}{\qty[(\omega-xp)^2+x(1-x)p^2]^2}=\nonumber\\
    =&18\lambda\delta_{ab}\delta_{\alpha\beta}\int_0^1\dd{x}\int_{\abs{l}\in[\Lambda,\Lambda_0]}\frac{\dd{l}}{2\pi}\frac{-l+(1-x)p}{\qty[l^2+x(1-x)p^2]^2}=\nonumber\\
    =&18\lambda\delta_{ab}\delta_{\alpha\beta}\int_0^1\dd{x}\int_{\abs{l}\in[\Lambda,\Lambda_0]}\qty(-\frac{1}{l^3}+\frac{(1-x)p}{l^4}+O(p^2))=\nonumber\\
    =&\frac{3\lambda\delta_{ab}\delta_{\alpha\beta}}{\pi}p\qty(\frac{1}{\Lambda^3}-\frac{1}{\Lambda_0^3})\,,
\end{align}
which is exactly the same result as before.

Putting everything together we get,
\begin{align}
    \lambda_{(3)}(\Lambda)&=\sqrt{\frac{\Lambda_0}{\Lambda}}\sqrt{\frac{\lambda}{N}}\qty[1+\frac{2\lambda}{3\pi}\qty(\frac{1}{\Lambda^3}-\frac{1}{\Lambda_0^3})]\\
    \lambda_{(4)}(\Lambda)&=\frac{\Lambda_0}{\Lambda}\frac{\lambda}{N}\qty[1+\frac{4\lambda}{3\pi}\qty(\frac{1}{\Lambda^3}-\frac{1}{\Lambda_0^3})]\,,
\end{align}
which now preserves supersymmetry at the quantum level.

Therefore, we have found a regulator that indeed preserves supersymmetry at least at 1-loop level. However, the physical interpretation of this regulator is not at all clear, and it does not seem to be usable beyond perturbation theory. Nevertheless, it would be interesting to see if similar phenomena occur for other theories in higher dimensions. We will not pursue this further in this manuscript, leaving it to future work.
\subsection{RG with smooth regulators} \label{subsec:smoothrg}
As we mentioned in the introduction, the last calculation was mostly a warm-up calculation before doing full quantum RG. However, in order to have a local notion of scale we cannot impose a cutoff in Fourier space. Indeed, if the cutoff depends on spacetime, the Fourier transform is no longer invertible\footnote{This is quite easy to see. For example, take some function $f(x)$, the normal Fourier transform with a cutoff would give you $f(x)=\displaystyle\int_{-\Lambda}^\Lambda\frac{\dd{k}}{2\pi}\ee^{\ii kx}\hat{f}(k)$, and we can easily check that indeed this is invertible: $\hat{f}(k)=\displaystyle\int_\mathbb{R}\dd{x}\ee^{-\ii kx}f(x)=\displaystyle\int_\mathbb{R}\dd{x}\ee^{-\ii kx}\displaystyle\int_{-\Lambda}^\Lambda\frac{\dd{q}}{2\pi}\ee^{\ii qx}\hat{f}(q)=\displaystyle\int_{-\Lambda}^\Lambda\frac{\dd{q}}{2\pi}\hat{f}(q)\displaystyle\int_\mathbb{R}\dd{x}\ee^{\ii (q-k)x}=\hat{f}(k)$. However, if we promote $\Lambda\to\Lambda(x)$ in the first step then we can't swap the order of the two integrals and therefore we can't invert the transformation.}. Therefore we need to use a smoother procedure. To that effect, we will use some basic exact RG technology to implement a smooth cutoff. We shall remain in momentum space for this section for convenience, in section \ref{subsec:smoothpos} we address how to extend this to position space. We only need the most basic ideas of exact RG, nonetheless, we will review them for completeness. We closely follow the derivation in the beginning of \cite{2009:Igarashi}, for some other reviews on the topic of exact RG you can refer to \cite{1998:Morris, 2001:Bagnuls, 2002:Berges, 2003:Polonyi,2012:Rosten}. 

Let us consider scalar field theory for illustration. The key idea is to introduce a function $K(x)$ such that:
\begin{enumerate}
    \item $K(x)$ is a smooth, non-increasing, positive function of $x$\,;
    \item $K(x)=1$ for $x<1$\,;
    \item $K(x)\to0$ as $x\to\infty$ sufficiently fast\,.
\end{enumerate}

\begin{figure}[h]
    \centering
    \includegraphics[scale=0.4]{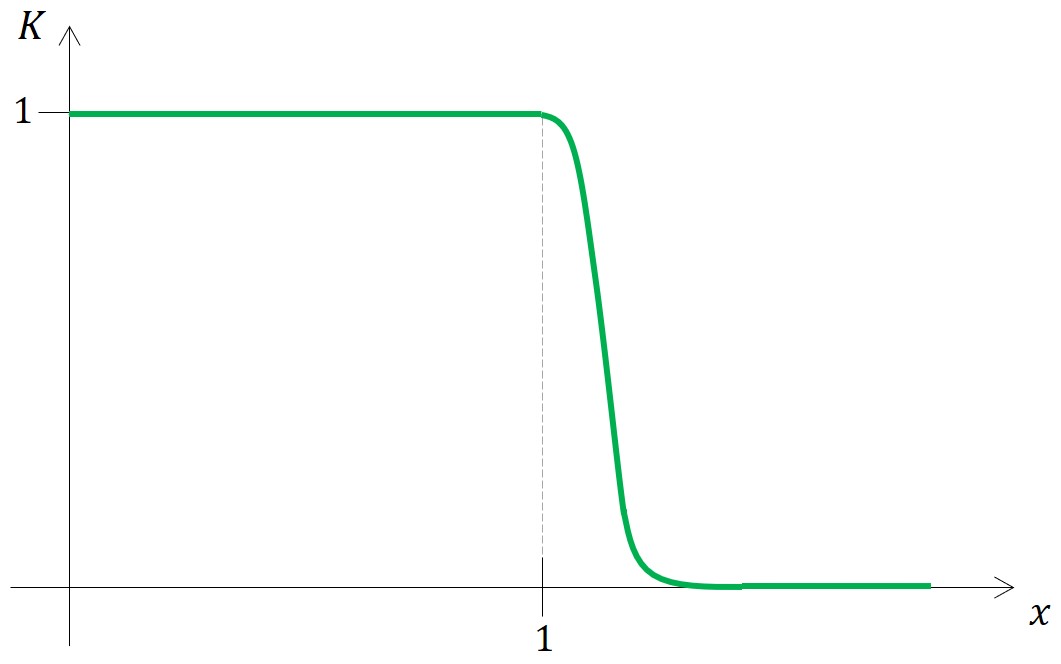}
    \caption{\label{fig:example}Example of an appropriate $K(x)$}
\end{figure}

See for example Fig~(\ref{fig:example}) for a function satisfying all the above criteria.

These requirements can be satisfied by a smooth function, however, no analytic function works. Nevertheless, we can soften the second requirement, and only impose that $K(0)=1$ and that $K(x)$ is suitably close to 1 for $x<1$. Then, we can find suitable analytic functions, \emph{e.g.} $K(x)=\ee^{-x^2}$. In momentum space, this distinction is not necessary, as there is no issue with working with smooth but non-analytic functions. However, when we go to position space, we need to phrase these functions in terms of operators, and therefore we need them to be analytic in order to be able to define them. With that in mind we shall assume we are using \textit{analytic} $K$, and therefore, Taylor expansions work.

The regulated action (with a global cutoff) is
\begin{equation}
    S_{\Lambda_0}=\frac{1}{2}\int_p \frac{p^2+m^2}{K\qty(\frac{p}{\Lambda_0})}\phi(p)\phi(-p)+S_\text{int}[\phi]\,,
\end{equation}
where, for brevity, we defined $\int_p\equiv\int\frac{\dd[d]{p}}{(2\pi)^d}$. 

Now, using the identity \cite{2009:Igarashi}
\begin{subequations}
\begin{align}
    &\int\mathcal{D}\phi_1~\mathcal{D}\phi_2~\exp[-\frac{1}{2}\int_p\frac{1}{A(p)}\phi_1(p)\phi_2(-p)-\frac{1}{2}\int_p\frac{1}{B(p)}\phi_2(p)\phi_2(-p)+S_\text{int}[\phi_1+\phi_2]]=\nonumber\\
    =&\left(\int\mathcal{D}\phi~\exp[-\frac{1}{2}\int_p\frac{1}{A(p)+B(p)}\phi(p)\phi(-p)-S_\text{int}[\phi]]\right)\times\nonumber
    \\
    &\left(\int\mathcal{D}\phi'~\exp[\frac{1}{2}\int_p\qty(\frac{1}{A(p)}+\frac{1}{B(p)})\phi'(p)\phi'(-p)]\right)\,,
    \label{eq:ergmaster}
\end{align}
where
\begin{align}
    \phi&=\phi_1+\phi_2\label{eq:ergbasis1}\\
    \phi'&=-\frac{B}{A+B}\phi_1+\frac{A}{A+B}\phi_2\label{eq:ergbasis2}\,,
\end{align}
we can write (by appropriately choosing $A$ and $B$, and neglecting the $\phi'$ integral since it only contributes with a field independent constant)
\begin{align}
    \int\mathcal{D}\phi~\ee^{-S_{\Lambda_0}[\phi]}=\int\mathcal{D}\phi_l\mathcal{D}\phi_h&\exp\Bigg[-\frac{1}{2}\int_p\frac{p^2+m^2}{K\qty(\frac{p}{\Lambda})}\phi_l(p)\phi_l(-p)-\nonumber\\
    &-\frac{1}{2}\int_p\frac{p^2+m^2}{K\qty(\frac{p}{\Lambda_0})-K\qty(\frac{p}{\Lambda})}\phi_h(p)\phi_h(-p)-S_\text{int}[\phi_h+\phi_l]\Bigg]\,,
\end{align}
\end{subequations}
which gives the required split into high and low energy modes, but now through a smooth regulator. 

The key point is that, when we are integrating over the high energy modes, the propagator can be approximated via,
\begin{equation}
    \frac{K\qty(\frac{p}{\Lambda_0})-K\qty(\frac{p}{\Lambda})}{p^2+m^2}=-\frac{pK'\qty(\frac{p}{\Lambda_0})\var{\Lambda}}{\Lambda_0^2}\frac{1}{p^2+m^2}+\mathcal{O}(\var{\Lambda}^2)
\end{equation}
for $\Lambda=\Lambda_0-\var{\Lambda}$.

This means that, if we are only interested in the beta functions, we only need to consider diagrams with one high energy propagator. Working with a smoother cutoff implies we count propagators instead of loops. Even if we are not just interested in the beta function and we want the full RG, this is still a relevant phenomenon. The analyticity of $K$ mean we can Taylor expand and compute the integrals order by order and different orders will not mix. We must count propagators.

This is manifestly at odds with supersymmetry. Now we cannot cancel the mass term for the scalars since the two diagrams come at a different order in $\var{\Lambda}$. To counter that, we could try lowering the fermionic and scalar cutoffs at a different rate so that each scalar propagator counts as two fermionic propagators, making both terms contributing to the scalar propagator appear at the same order and allowing the mass term to cancel. However, even in that case, supersymmetry is broken. The reason now being that the corrections to the other couplings come at higher orders in $\var{\Lambda}$, so the only contribution to the beta function would be from the scalar wavefunction renormalisation, and there is one scalar for the cubic coupling, but four scalars in the quartic coupling. We would not have $\lambda_{(3)}^2=\lambda_{(4)}$ and supersymmetry would be broken.

By itself this is not a very surprising result, a similar phenomenon already happens for the much simpler four-dimensional $\mathcal{N}=1$ theory with one complex scalar and one Weyl fermion. In this case, however, one can preserve supersymmetry, even with a smooth regulator, by using the off-shell formalism. This is accomplished by using auxiliary fields that make the supersymmetry algebra close without using the equations of motion. This was our issue previously, by introducing a regulator in the style described above, we have changed the equations of motion, which were essential in preserving supersymmetry. Then, if we regulate all quadratic terms with the same function, including the auxiliary field, which now becomes dynamical and propagating, we do not break supersymmetry. This can happen because we no longer have the quartic scalar coupling, what we do have is a cubic coupling with two scalars and one auxiliary field. This means (using dotted lines for auxiliary fields),

\begin{equation}
\begin{tikzpicture}[baseline=(b)]
  \begin{feynman}
    \vertex (a);
    \vertex [right=1.5cm of a] (centre);
    \vertex [right=1.5cm of centre] (b);
    \vertex [above=1.5cm of centre] (aux);
    
    \diagram* {
      (a) -- [scalar,out=0,in=180] (centre),
      (centre) -- [scalar,out=0,in=180] (b),
      (centre) -- [blue,scalar,out=135,in=180] (aux),
      (aux) -- [blue,scalar,out=0,in=45] (centre),
    };
    \draw[fill=black] (centre) circle (1.5pt);
  \end{feynman}
\end{tikzpicture}
\to
\begin{tikzpicture}[baseline=(a)]
  \begin{feynman}
    \vertex (a);
    \vertex [right=1.4cm of a] (centre1);
    \vertex [right=0.6cm of centre1] (aux1);
    \vertex [above=0.6cm of aux1] (aux2);
    \vertex [below=0.6cm of aux1] (aux3);
    \vertex [right=0.6cm of aux1] (centre2);
    \vertex [right=1.4cm of centre2] (b);

    \diagram* {
      (a) -- [scalar] (centre1),
      (centre1) -- [blue,scalar,out=90,in=180] (aux2),
      (aux2) -- [blue,scalar,out=0,in=90] (centre2),
      (centre2) -- [blue,ghost,out=-90,in=0] (aux3),
      (aux3) -- [blue,ghost,out=180,in=-90] (centre1),
      (centre2) -- [scalar] (b),
    };
  \end{feynman}
  \draw[fill=black] (centre1) circle (1.5pt);
  \draw[fill=black] (centre2) circle (1.5pt);
\end{tikzpicture}
\end{equation}
which comes at the same order as the fermionic loop.

Knowing this result for the simpler theory, could we reproduce this with BFSS? The answer turns out to be no. Our first hurdle is the fact that no off-shell formulation with this many supercharges and finitely many fields is known\footnote{We thank Nick Dorey for pointing that out to us.}. We can try to ameliorate our situation by using the $\mathcal{N}=1$ superspace formulation of four-dimensional $\mathcal{N}=4$ SYM and dimensionally reducing it down to $1D$. In this manner we would have 4 supercharges preserved off-shell. However, this is still not enough to prevent the formation of a mass term. This happens because we do not destroy every quartic coupling, just some of them, so part of the calculation that leads to the mass term would carry through with no change. To implement a smooth cutoff, which we must do to make it local, means giving up explicit supersymmetry.
\section{Local Renormalisation Group} \label{sec:lrg}
In this section we do the first step in performing QRG, defining how to integrate out modes with a local regulator, \emph{i.e.} integrating modes at different speeds in each point of spacetime. To do that we first repeat the derivation done in section \ref{subsec:smoothrg} but now in position space. We shall see that it still holds, provided there are some restrictions on the kinetic operators we use. Then we take a particular example, of a local Gaussian regulator in one dimension and prove that that regulator obeys all necessary restrictions. This provides the first explicit realisation of a local cutoff scheme which could be used in practical calculations.
\subsection{Smooth regulator in position space} \label{subsec:smoothpos}
Deriving (\ref{eq:ergmaster}) is rather straightforward. As it stands, we just plug in the definitions (\ref{eq:ergbasis1}) and (\ref{eq:ergbasis2}) and do the resulting algebra. The reason for this simplicity is that, in momentum space, we are dealing with ordinary multiplication of functions. In position space, however, we would be dealing with operators. Which do not obey many of the nice properties we take for granted when performing algebraic manipulations.

For simplicity, we shall resort to matrix multiplication notation, where spacetime integration is denoted with a dot product. In this notation, local operators become matrices by introducing a delta function\footnote{Throughout this paper we always work in Euclidean signature.},

\begin{equation}
    -\int\dd[d]x\phi(x)\lap{\phi(x)}=-\int\dd[d]{x}\dd[d]{y}\phi(x)\qty[\lap{\delta^{(d)}(x-y)}]\phi(y)=\phi^\TT G^{-1}\phi
\end{equation}
where 
\begin{equation}
    G^{-1}(x,y)=-\lap{\delta^{(d)}(x-y)}
    \label{eq:example}
\end{equation}
and, as usual, the inverse of the operator will be it's Green's function. For example
\begin{equation}
    G(x,y)=\int\frac{\dd[d]k}{(2\pi)^d}\frac{1}{k^2}\ee^{\ii k\vdot(x-y)}\,,
    \label{eq:green}
\end{equation}
so that,
\begin{equation}
    G^{-1}G=\int\dd[d]{z}G^{-1}(x,z)G(z,y)=-\nabla^2_xG(x,y)=\delta^{(d)}(x-y)=\mathbb{1}\,.
\end{equation}

It is important to note that, in general, these objects will not obey the same nice properties that matrices do. Namely, for a given "matrix" (\emph{i.e.} function of two arguments), left and right inverses do not necessarily match, and the inverse of a diagonal object is not necessarily diagonal. Note, for instance, that (\ref{eq:example}) is diagonal, but (\ref{eq:green}) is not. In this case left and right inverses do match because both are symmetric.  In the end these subtleties will not be all that relevant, but it is important to have in mind the full picture.

Let us start by deriving (\ref{eq:ergmaster}) in position space. We take $B^{-1}=G_{\Lambda}^{-1}$ to be the low energy kinetic operator and $(A+B)^{-1}=G_{\Lambda_0}^{-1}$ to be the high energy kinetic operator. We make no assumption at this point as to whether they are local or global regulators. However, by construction they will both be symmetric. So, if the inverses exist, they will behave as expected. 

If we can find those two inverses, $B$ and $A+B$, we can define $A=A+B-B=G_{\Lambda_0}-G_\Lambda$ as the high energy propagator, which is the most useful quantity in practical calculations, and, by construction, is also symmetric. Then, if $A^{-1}$ exists, it behaves just like a matrix inverse. Just note that finding $A^{-1}$ can be incredibly hard because it is the opposite question to what is usually done, we have the Green's function and we want to find out the corresponding operator. However, even though our derivation will only work if such an operator actually exists, we do not actually need for any practical calculations so it suffices to show that it exists.

We repeat the derivation of (\ref{eq:ergmaster}) assuming all those inverses behave as expected, and, in the next section, we present an explicit example and check whether these assumptions are valid. We will be careful in saying exactly what conditions are needed, so that, in future work it is clear if any generalisation is possible.

Analogously to (\ref{eq:ergbasis1}) and (\ref{eq:ergbasis2}), we start by defining:
\begin{subequations}
\begin{align}
    \phi_h&=A(A+B)^{-1}\phi-\phi'
    \\
    \phi_l&=B(A+B)^{-1}\phi+\phi'\,,
\end{align}
\end{subequations}
so that the Jacobian is still unity.

We therefore have (ignoring an overall, unimportant, factor of $\frac{1}{2}$):

\begin{align}
    &\phi_h^\TT A^{-1}\phi_h+\phi_l^\TT B^{-1}\phi_l=\nonumber\\
    =&\phi^\TT\qty[(A+B)^{-1}]^\TT A^\TT A^{-1}A(A+B)^{-1}\phi+\phi'^\TT A^{-1}\phi'-\nonumber\\
    -&\phi^\TT\qty[(A+B)^{-1}]^\TT A^\TT A^{-1}\phi'-\phi'^\TT A^{-1}A(A+B)^{-1}\phi+\nonumber\\
    +&\phi^\TT\qty[(A+B)^{-1}]^\TT B^\TT B^{-1}B(A+B)^{-1}\phi+\phi'^\TT B^{-1}\phi'+\nonumber\\
    +&\phi^\TT\qty[(A+B)^{-1}]^\TT B^\TT B^{-1}\phi'+\phi'^\TT B^{-1}B(A+B)^{-1}\phi\,.
\end{align}
Using, $B^\TT B^{-1}=\mathbb{1}$, $A^{-1}A=\mathbb{1}$, $\qty[(A+B)^{-1}]^\TT(A^\TT+B^\TT)=\mathbb{1}$, and $A^\TT A^{-1}=\mathbb{1}$ we find,
\begin{equation}
    \phi^\TT(A+B)^{-1}\phi+\phi'^\TT(A^{-1}+B^{-1})\phi'
\end{equation}
which is the desired expression.

Note that if all operators and Green's functions are symmetric, which implies left and right inverses match, then all conditions are satisfied. Also note that we are free to choose both the high energy and the low energy propagators, so the real crux is on the properties of $A$ and the existence of $A^{-1}$.
\subsection{An example: local Gaussian regulator} \label{subsec:localgauss}
We shall restrict to Euclidean time and consider a Gaussian regulator. In the end we shall be most interested in the case $d=1$ but the results of this section are valid for arbitrary $d$. With a usual, spacetime independent cutoff we have,
\begin{subequations}
\begin{equation}
    G_{\Lambda}^{-1}(x_1,x_2)=-\ee^{-\frac{\lap_{x_2}}{\Lambda^2}}\lap_{x_2}\delta^{(d)}(x_1-x_2)
    \label{eq:globalop}
\end{equation}
which has the Green's function,

\begin{equation}
    G_{\Lambda}(x_1,x_2)=\int\frac{\dd[d]{k}}{(2\pi)^d}\frac{\ee^{-\frac{k^2}{\Lambda^2}}}{k^2}\ee^{\ii k\vdot(x_1-x_2)}\label{eq:lambda0fourier}
\end{equation}
such that the kinetic term looks like

\begin{equation}
    -\int\dd[d]{x}\phi(x)\ee^{-\frac{\lap}{\Lambda^2}}\lap\phi(x)\,.
    \label{eq:globalkinetic}
\end{equation}
\end{subequations}
In what follows we need to give the cutoff spacetime dependence. If we naively just promote $\Lambda\to\Lambda(x)$ directly in (\ref{eq:globalkinetic}) there will be ordering issues when expanding the exponential which will make it hard to deal with. To help with that, we start with (\ref{eq:globalop}) instead and promote $\Lambda\to\Lambda(x_1)$. In this way the derivatives actually commute with the cutoff, so there are no ordering issues. However, then the resulting operator is not symmetric (and only the symmetric part contributes to the action because it's multiplied on both sides by the same field). Therefore, we take the symmetric part and define the local version as,
\begin{equation}
    G_{\Lambda}^{-1}(x_1,x_2)=-\frac{1}{2}\qty(\ee^{-\frac{\lap_{x_2}}{\Lambda(x_1)^2}}\lap_{x_2}+\ee^{-\frac{\lap_{x_1}}{\Lambda(x_2)^2}}\lap_{x_1})\delta^{(d)}(x_1-x_2)\,.
    \label{eq:localop}
\end{equation}

Unfortunately, for arbitrary $\Lambda(x)$ we do not know how to find the Green's function of (\ref{eq:localop}). However, for our purposes (as will be shown in the following section), we only need to find the beta functions, \emph{i.e.} infinitesimal flow. Therefore we approximate, defining the original high energy cutoff, $\Lambda_0$ to be constant, and taking, $\Lambda(x)=\Lambda_0\ee^{-\alpha(x)dz}$, for $\alpha$ and $dz$ positive, and $dz\ll1$. We can then solve this perturbatively, expanding in powers of $dz$,

\begin{align}
    G_{\Lambda}^{-1}(x_1,x_2)=&-\ee^{-\frac{\lap_{x_2}}{\Lambda_0^2}}\lap_{x_2}\delta^{(d)}(x_1-x_2)+\nonumber\\
    &+\frac{dz}{\Lambda_0^2}\qty(\alpha(x_1)\ee^{-\frac{\lap_{x_2}}{\Lambda_0^2}}\nabla_{x_2}^4+\alpha(x_2)\ee^{-\frac{\lap_{x_1}}{\Lambda_0^2}}\nabla_{x_1}^4)\delta^{(d)}(x_1-x_2)+\mathcal{O}(dz^2)\,.
    \label{eq:localpert}
\end{align}

This all means we have chosen $G_{\Lambda_0}^{-1}=(A+B)^{-1}$, $G_{\Lambda}^{-1}=B^{-1}$, for $\Lambda_0$ constant as in (\ref{eq:globalop}) and $\Lambda(x)=\Lambda_0\ee^{-\alpha(x)dz}$. We then expand,
\begin{equation}
    G_\Lambda(x_1,x_2)=G_{\Lambda_0}(x_1,x_2)+dzG^{(1)}(x_1,x_2)+\mathcal{O}(dz^2)
\end{equation}
giving us $A=G_{\Lambda_0}-G_\Lambda=-dz\,G^{(1)}$. All we have to do now is find $A$ and show that $A^{-1}$ exists.

First we find $A$, \emph{i.e.} the Green's function for (\ref{eq:localpert}), order by order in powers of $dz$. At 0$^\text{th}$ order, the equation is solved by construction. At 1$^\text{st}$ order we get,

\begin{align}
    -\ee^{-\frac{\lap_{x_1}}{\Lambda_0^2}}\lap_{x_1}G^{(1)}(x_1,x_2)=&-\frac{\alpha(x_1)}{\Lambda_0^2}\ee^{-\frac{\lap_{x_1}}{\Lambda_0^2}}\nabla_{x_1}^4G_{\Lambda_0}(x_1,x_2)-\frac{1}{\Lambda_0^2}\ee^{-\frac{\lap_{x_1}}{\Lambda_0^2}}\nabla_{x_1}^4\qty(\alpha(x_1)G_{\Lambda_0}(x_1,x_2))\\
\intertext{Using the definition of $G_{\Lambda_0}$ as the Green's function for (\ref{eq:globalop}) allows us to simplify the first term on the RHS,}
    -\ee^{-\frac{\lap_{x_1}}{\Lambda_0^2}}\lap_{x_1}G^{(1)}(x_1,x_2)=&\frac{\alpha(x_1)}{\Lambda_0^2}\lap_{x_1}\delta^{(d)}(x_1-x_2)-\frac{1}{\Lambda_0^2}\ee^{-\frac{\lap_{x_1}}{\Lambda_0^2}}\nabla_{x_1}^4\qty(\alpha(x_1)G_{\Lambda_0}(x_1,x_2))\\
\intertext{Acting with $G_{\Lambda_0}$ on the left on both sides of this equation, and, once more using its defining property as the Green's function gives us,}
    G^{(1)}(x_1,x_2)=&\frac{1}{\Lambda_0^2}\qty(\lap_{x_2}\qty(G_{\Lambda_0}(x_1,x_2)\alpha(x_2))+\lap_{x_1}\qty(\alpha(x_1)G_{\Lambda_0}(x_1,x_2)))\,.
    \label{eq:uvprop}
\end{align}
Everything is nice and symmetric as expected, which means left and right inverses will match nicely, if they do exist, that is. As mentioned above we do not actually need to find an explicit expression for the inverse, we just need to know that it indeed exists to render our calculations consistent.

It is instructive to take the Fourier transform of (\ref{eq:uvprop}), using the explicit expression in (\ref{eq:lambda0fourier}). After a straightforward calculation, just using the definition of Fourier transform and some manipulation of delta-functions we arrive at,
\begin{equation}
    \hat{G}^{(1)}(k_1,k_2)=-\frac{1}{\Lambda_0^2}\qty(\frac{k_2^2}{k_1^2}\ee^{-\frac{k_1^2}{\Lambda_0^2}}+\frac{k_1^2}{k_2^2}\ee^{-\frac{k_2^2}{\Lambda_0^2}})\hat{\alpha}(k_1+k_2)\,.
    \label{eq:fourierg1}
\end{equation}
Because everything is nice and symmetric, left and right inverses match, and we can then use standard linear algebra results. In this language, an inverse exists if and only if
\begin{equation}
    \int\frac{\dd[d]{k_2}}{(2\pi)^d}\hat{G}^{(1)}(k_1,k_2)f(-k_2)=0~~\forall k_1\Rightarrow f(k_2)=0~~\forall k_2\,,
    \label{eq:invcond}
\end{equation}
where, crucially, $f$ cannot have any dependence on $k_1$.

Imagine for a moment that in (\ref{eq:fourierg1}) there is no $\hat{\alpha}$, then this is clearly not true. We just need to pick $f$ to be an odd function and the integral vanishes. This is also the case for constant $\hat{\alpha}$, however, a constant $\hat{\alpha}$ corresponds to a delta-function in position space, which we can clearly rule out as an allowed profile for $\hat{\alpha}$, it would correspond to changing the scale only at one point. So let us restrict to the case when $\hat{\alpha}$ is not constant.

In this case for a given $\hat{\alpha}$, and a given $k_1$, we could conceivably make the integral vanish for a non-zero $f$ by judiciously choosing $f$, possibly relying on some non-trivial symmetry. However, because $\hat{\alpha}$ only depends on the combined sum $k_1+k_2$, any such choice will inevitably depend on $k_1$. Unless $\hat{\alpha}$ is constant (which we have ruled out), by just choosing a different $k_1$ we will shift the profile of $\hat{\alpha}$ in an arbitrary fashion, and inevitably, some of those shifts will ruin our choice of $f$. Given that $f$ cannot depend on $k_1$ and the condition must be valid for all $k_1$ we conclude that (\ref{eq:invcond}) is true, and therefore, $G^{(1)}$ is invertible, rendering our procedure consistent. We have successfully developed an RG scheme with a local change of scale.
\section{Quantum Renormalisation Group}\label{sec:qrg}
After developing a framework to perform local RG we can move on to the main objective of this paper, testing quantum RG (QRG). We start by an overview of the procedure itself in greater detail than what was given in the introduction. We then move on to an overview of what is known (and is relevant) about the holographic duality of the BFSS model, to understand what should be our starting point and are we expecting to reproduce (or fail to reproduce) after performing QRG. Finally we put everything together and do the actual computation.
\subsection{Overview of QRG}\label{sec:overviewqrg}
The starting point for QRG \cite{2010:Lee,2012:Lee,2014:Lee} is a quantum field theory with dynamical fields $\Phi$ which are matrix valued. These could have any spin, but it is important that they are matrix valued. We write the partition function of this theory as

\begin{equation}
    Z=\int\mathcal{D}\Phi~\exp(\ii S_0[\Phi])
\end{equation}

The algorithm of QRG is as follows:

\begin{enumerate}
    \item Turn on single trace operator deformations. In general, we should turn on a complete basis of single trace operators. However, in practice, we will only be able to turn on a finite number of them. Let $O_m$ be the operators and $j^{(0)m}$ be the corresponding sources, the partition function is then
    \begin{subequations}
    \begin{equation}
        Z[j^{(0)}]=\int\mathcal{D}\Phi~\exp(\ii S_0[\Phi]+\ii S_1[O,j^{(0)}])
    \end{equation}
    where    
    \begin{equation}
        S_1[O,j^{(0)}]=N^2\sum_m\int\dd[d]{x}O_mj^{(0)m}\,.
    \end{equation}
    \end{subequations}    
    
    \item Perform an infinitesimal local change of scale, \emph{i.e.} if in the initial theory the cutoff is $\Lambda_0$, we do an RG flow such that the new scale is $\Lambda=\ee^{-\alpha^{(1)}(x) dz}\Lambda_0$, for $dz\ll1$. The new partition function is, to leading order in $dz$,
    \begin{subequations}    
    \begin{equation}
        Z[j^{(0)}]=\int\mathcal{D}\Phi~\exp(\ii S_0[\Phi]+\ii\var{S}[O,j^{(0)}]+\ii S_1[O,j^{(0)}])
    \end{equation}
    where    
    \begin{align}
        \var{S}[O,j^{(0)}]=dz N^2\int\dd[d]{x}\bigg\{&\mathcal{L}_\text{C}(x;j^{(0)}]-\beta^m(x;j^{(0)}]O_m+\nonumber\\
        &+\frac{1}{2}G^{mn\qty{\mu}}(x;j^{(0)}]O_m\partial_\qty{\mu}O_n\bigg\}
        \label{deltaS}
    \end{align}
    \end{subequations}
    and $f(x;j^{(0)}]$ denotes a function that depends on $j^{(0)}(x)$ and its derivatives at a point $x$. We have used the fact we turned on a complete basis of operators to write all appearances of the fields in terms of the operators we have turned on. If we only turn on a finite number of them, it cannot generate any new ones, or otherwise this is not a consistent algorithm. Note that, to leading order in $dz$, we do not generate more than double trace operators.
    
    \item Substitute $O_m\to-\frac{\ii}{N^2}\fdv{j^{(0)m}}$ in $\var{S}$, noting that we must now be careful with the ordering in (\ref{deltaS}), the order shown, where all the operators are on the right is the correct one
    
    \begin{equation}
        Z[j^{(0)}]=\int\mathcal{D}\Phi~\exp(\ii S_0[\Phi]+\ii\var{S}\qty[-\frac{\ii}{N^2}\fdv{j^{(0)}},j^{(0)}]+\ii S_1[O,j^{(0)}])\,.
    \end{equation}
    
    \item Add auxiliary fields $p^{(1)}_m$ and $j^{(1)m}$ such that 
    
    \begin{align}
        Z[j^{(0)}]=\int&\mathcal{D}\Phi\prod_n\mathcal{D}p^{(1)}_n\mathcal{D}j^{(1)n}~\exp\bigg(\ii N^2\int\dd[d]{x}\sum_mp^{(1)}_m(j^{(1)m}-j^{(0)m})+\nonumber\\
        &+\ii S_0[\Phi]+\ii\var{S}\qty[-\frac{\ii}{N^2}\fdv{j^{(1)}},j^{(0)}]+\ii S_1[O,j^{(1)}]\bigg)\,.
    \end{align}
    
    \item Integrate by parts with respect to $j^{(1)m}$ in the $\var{S}$ term
    
    \begin{align}
        Z[j^{(0)}]=\int\prod_n\mathcal{D}p^{(1)}_n\mathcal{D}j^{(1)n}~\exp\bigg(&\ii N^2\int\dd[d]{x}\sum_mp^{(1)}_m(j^{(1)m}-j^{(0)m})+\nonumber\\
        &+\ii\var{S}\qty[-p^{(1)},j^{(0)}]\bigg)Z[j^{(1)}]\,.
    \end{align}
    
    \item Now we can start with $Z[j^{(1)}]$ and iterate this procedure
    
    \begin{align}
        Z[j^{(0)}]=\int\prod_{l=1}^L\prod_n\mathcal{D}p^{(l)}_n\mathcal{D}j^{(l)n}~\exp\Bigg(&\ii N^2dz\sum_{l=1}^L\int\dd[d]{x}\sum_mp^{(1)}_m\qty(\frac{j^{(l)m}-j^{(l-1)m}}{dz})+\nonumber\\
        &+\ii\sum_{l=1}^L\var{S}\qty[-p^{(l)},j^{(l-1)}]\Bigg)Z[j^{(l)}]\,.
    \end{align}
    
    Taking the $dz\to0$ limit, it's not hard to see we have generated an action that lives in $d+1$ dimensions for the new dynamical fields $j^m(z,x)$ and $p^m(z,x)$. It is important to note that if no double trace operators are generated then this action will be linear in $p^m(z,x)$ and therefore this field will still just be a Lagrange multiplier, not a dynamical field. In order to have non-trivial dynamics for these fields we must generate double trace operators.
\end{enumerate}

The conjecture is that Gauge/Gravity Duality is completely encapsulated in a procedure such as this one. As mentioned in the introduction, there has been some additional work with regards to this conjecture. Namely some hints for it's application to the original AdS$_5$/CFT$_4$ case \cite{2015:Nakayama}, a concrete calculation for the $U(N)$ vector model \cite{2015:Lunts}, and, understanding the conditions under which one can recover full $(d+1)$-dim diffeomorphism invariance \cite{2017:Shyam}. This last one is the most relevant for our purposes since it his here that the importance of having a spacetime dependent cutoff was fully appreciated as a means to recover diffeomorphism invariance.
\subsection{Overview of the holographic dual to BFSS}
As mentioned in section \ref{subsec:reviewbfss}, the BFSS model, describes the dynamics of $N$ coincident D0-branes. This means it also has a dual gravitational description in terms of 10-dim type IIA supergravity \cite{1998:Itzhaki}. In the decoupling limit,

\begin{equation}
    U=\frac{r}{\alpha'}=\text{fixed},~g_\text{YM}^2=\frac{1}{4\pi^2}\frac{g_s}{\alpha'^{3/2}}=\text{fixed},~\alpha'\to0
\end{equation}
the supergravity background solution corresponding to BFSS is given by \cite{1998:Itzhaki},
\begin{subequations}
\begin{align}
    \mathrm{d}s^2&=\alpha'\qty(-\frac{U^{7/2}}{4\pi^2g_\text{YM}\sqrt{15\pi N}}\dd{t}^2+\frac{4\pi^2g_\text{YM}\sqrt{15\pi N}}{U^{7/2}}\dd{U}^2+\frac{4\pi^2g_\text{YM}\sqrt{15\pi N}}{U^{3/2}}\mathrm{d}\Omega^2)\,,\label{eq:back}
    \\
    \nonumber
   \\
    \ee^\phi&=4\pi^2g_\text{YM}^2\qty(\frac{240\pi^5g_{YM}^2N}{U^7})^{3/4}\,.
\end{align}
\end{subequations}
where $\mathrm{d}\Omega^2$ is the metric on a round unit radius $S^8$, $\alpha^\prime$ is related to the string length and $g_s$ is the string coupling. We note in passing that strictly speaking this solution is singular at the origin. The standard way to deal with this is to put the system at a finite temperature, which corresponds to having a black hole in the gravity perspective. However, if we are far enough away from the origin, \emph{i.e.} near the boundary, the effects of this temperature should be minimal, that is also the region we are we have more control over our field theoretic description. Therefore, in this paper, we neglect finite temperature effects.

Another point to make is that, as mentioned in \cite{1998:Itzhaki}, the curvature gets large as we approach the boundary, more specifically, $\alpha' R\sim\sqrt{\frac{U^3}{g_\text{YM}^2N}}$, and therefore we have less faith on our supergravity description. Naively, this does not intersect with the region where we have analytic control on the field theory side. However, in QRG we only need to do one infinitesimal step of coarse graining, and, as we have showcased in sections \ref{subsec:smoothrg} and \ref{subsec:localgauss} we can do that exactly. This seems to solve all our problems, but there is an issue. The theory we want is the one that approaches the action \ref{eq:bfssaction} in the UV. When the coupling is strong, the correct action is not \ref{eq:bfssaction} but it needs corrections that, by construction, will be very important. If we just take the action \ref{eq:bfssaction} and define the coupling to be strong we have a well defined theory and calculations, it will simply not be the theory we're looking after. This is similar to how we can solve QCD in the strong coupling limit exactly using lattice methods but the answers we get aren't physically relevant\footnote{We thank David Tong for pointing this out.}.

The resolution to this issue comes from the realisation that, in the gravity side, we should insert the sources at the boundary, not deep into the bulk. Therefore, we should start with a field theory action in the $\Lambda_0\to\infty$ limit. Then we do the one infinitesimal coarse graining step required by QRG at this weak coupling limit. By the nature of QRG we can put all corrections due to this step into the new dynamical fields and start again with the original action. This means we can confidently do all the hard calculations in the regime where we have control over the theory and then use the auxiliary turned dynamical fields to recover the important physics.


With those points in mind we carry on with our discussion. The solution presented above is not the full content of the gauge/gravity duality. As was mentioned in the introduction, the most general form of the correspondence is an equality between partition functions that allows us to calculate correlation functions on both sides (and hopefully match them) \cite{1999:Maldacena,1998:Gubser,1998:Witten,2000:Aharony}. However, to do that, we need to find out which operators on the field theory side correspond to which modes on the gravity side.

This is precisely what was done in \cite{2000:Sekino}. By decomposing the ten-dimensional modes in harmonics of the eight-dimensional sphere, they have found a correspondence between certain supergravity modes and certain operators discussed in \cite{1999:Taylor}. In addition to harmonic analysis, a very important tool is generalised conformal symmetry, which, despite its importance, is not very pertinent to the main point of this paper, so we skip it, for interested readers here is a selection of useful literature on the subject \cite{1998:Jevicki,1999:Jevicki,1999:Hayasaka,1999:Hata,2008:Kanitscheider,2018:Taylor}.

We will not repeat here the full dictionary except to point out that these modes are constructed such that, up to quadratic order in the supergravity action, they do not mix and have an effective two-dimensional action. Therefore, if we turn on the correct operator in the field theory side, even if just that one, we should be capable of reproducing the correct 2-point function on the gravity side. This test has indeed been made in \cite{2011:Hanada} and matching between the two sides has been found.

In particular we shall turn on the operator \cite{1999:Taylor}
\begin{equation}
    T^{++}_{2,ij}=\frac{1}{\lambda^{9/7}N}\Tr(X_iX_j-\frac{\delta_{ij}}{9}X_kX_k)
    \label{eq:op}
\end{equation}
which is dual to the supergravity mode \cite{2000:Sekino}
\begin{equation}
    s_3^{\ell=2}=z^{-7/5}\qty(-7b^i_i+f)
\end{equation}
where
\begin{subequations}
\allowdisplaybreaks
\begin{align}
    z&=\frac{2}{5}q^{1/2}r^{-5/2}\\
    q&=60\pi^3(\alpha')^{7/2}g_sN\\
    f&=g_s\qty(\frac{5}{2})^{19/5}q^{-2/5}z^{19/5}(\partial_0 a_z-\partial_z a_0)\\
    h^i_i(x^\mu)&=\sum b^i_i(t,z)Y(x^i)\\
    \hat{A}_z(x^\mu)&=\sum a_z(t,z)Y(x^i)\\
    \hat{A}_0(x^\mu)&=\sum a_0(t,z)Y(x^i)\,,
\end{align}
\end{subequations}
and $Y$ are the scalar $SO(9)$ spherical harmonics (we have suppressed their internal indices), $h_{\mu\nu}$, $\hat{A}_\mu$ are the perturbations of, respectively, the metric and the gauge field around the background (\ref{eq:back}).

These modes, have the following 2-point function, as discussed in \cite{2000:Sekino} and confirmed in \cite{2011:Hanada}:
\begin{equation}
    \expval{\mathcal{O}(\tau)\mathcal{O}(\tau')}_c=\frac{1}{\kappa^2}q^{29/35}\frac{1}{\abs{\tau-\tau'}^{-1/5}}
\end{equation}
which we should be able to reproduce if QRG is valid. We shall assess in the following whether or not QRG holds.

Finally, a quick note that, for this simple case, there is the possibility of recovering interactions because there are known fully consistent truncations down to 2 dimensions \cite{2013:Ortiz,2013:Anabalon,2014:Ortiz}, which do agree with the tests performed in \cite{2011:Hanada}. Even though we shall use the fact such truncations exist to draw some conclusions we shall not need to use the particular structure therefore we refer the reader to the above cited literature.
\subsection{QRG of BFSS}
We now apply the full QRG calculation of the action (\ref{eq:bfssaction}). We first add the source term,
\begin{equation}
    N^2\int\dd{\tau}J^{++}_{2,ij}(\tau)T^{++}_{2,ij}(\tau)
\end{equation}
where $T^{++}_{2,ij}$ is given by (\ref{eq:op}).

In this case, because it is very important to keep track of the trace structure we shall resort to fundamental indices $I,J=1,\dots,N$, and represent the fields by traceless Hermitian matrices. It is also simpler to use Wick's theorem instead of Feynman diagrams. We shall furthermore be agnostic about the regulator procedure used, just noting that is has to be local, and could be, for example, the one developed in section \ref{subsec:localgauss} (it is not too hard to generalise the results of that section to include fermions). We then split all the index structure apart from the temporal dependence and write, for the high energy modes to be integrated out,
\begin{subequations}
\begin{align}
    \expval{\tilde{X}^{+I}_{i~J}(\tau_1)\tilde{X}^{+K}_{j~L}(\tau_2)}_+&=A_X(\tau_1,\tau_2)\delta_{ij}\qty(\delta^I_L\delta^K_J-\frac{1}{N}\delta^I_J\delta^K_L)\label{eq:scalarprop}\\
    \expval{\tilde{\psi}^{+I}_{\alpha J}(\tau_1)\tilde{\psi}^{+K}_{\beta L}(\tau_2)}_+&=A_\psi(\tau_1,\tau_2)\delta_{\alpha\beta}\qty(\delta^I_L\delta^K_J-\frac{1}{N}\delta^I_J\delta^K_L)\label{eq:fermiprop}
\end{align}
\end{subequations}
where we are using the rescaled variables defined in (\ref{eq:rescale}).

All we have to do is compute all connected correlation functions with just a single contraction, \emph{i.e.} propagator. Up to that order, the only terms that contribute are those that come from the expectation value of a single operator, or from the expectation value of the product of two operators. All such calculations proceed in exactly the same manner: expand the expectation value; pick all possible pairs of fields to be "+", \emph{i.e.} high energy, summing over all possible choices, the remaining fields become "-"; (anti-)commute past each other (depending if they're scalars or fermions) until you have expressions of the form (\ref{eq:scalarprop}) or (\ref{eq:fermiprop}); contract all indices noting that $\delta^I_I=N$. Therefore, we shall only present the full details for the first calculation and for all others we merely give the final answer. We note, however, that everything that involves the quartic interaction is much more cumbursome than anything else, because we need to sum over the possible choices.
\subsubsection{Single operator:}
\paragraph{Cubic interaction:}

\begin{align}
    &\int\dd{\tau}\expval{\frac{1}{2}\sqrt{\frac{\lambda}{N}}\Tr{\tilde{\psi}_\alpha(\gamma_i)_{\alpha\beta}\comm{\tilde{X}_i}{\tilde{\psi}_\beta}}}_+=\nonumber\\
    =&\int\dd{\tau}(\gamma_i)_{\alpha\beta}\frac{1}{2}\sqrt{\frac{\lambda}{N}}\Tr{\expval{\tilde{\psi}^+_\alpha\comm{\tilde{X}^-_i}{\tilde{\psi}^+_\beta}}}_+=\nonumber\\
    =&\int\dd{\tau}(\gamma_i)_{\alpha\beta}\frac{1}{2}\sqrt{\frac{\lambda}{N}}\qty(\expval{\tilde{\psi}^{+I}_{\beta J}\tilde{\psi}^{+J}_{\alpha K}}_+\tilde{X}^{-K}_{i~I}-\expval{\tilde{\psi}^{+I}_{\alpha J}\tilde{\psi}^{+J}_{\beta K}}_+\tilde{X}^{-K}_{i~I})=\nonumber\\
    =&\int\dd{\tau}(\gamma_i)_{\alpha\beta}\frac{1}{2}\sqrt{\frac{\lambda}{N}}A_\psi(\tau,\tau)\delta_{\alpha\beta}\qty(\qty(\delta^I_KN-\frac{1}{N}\delta^I_K)-\qty(\delta^I_KN-\frac{1}{N}\delta^I_K))\tilde{X}^{-K}_{i~I}=0
\end{align}

\paragraph{Quartic interaction:}

\begin{equation}
    \int\dd{\tau}\expval{-\frac{\lambda}{4N}\Tr{\comm{\tilde{X_i}}{\tilde{X_j}}^2}}_+=8\lambda\int\dd{\tau}A_X(\tau,\tau)\Tr{(\tilde{X}^-_i)^2}
\end{equation}

\paragraph{Source term:}

\begin{equation}
    N^2\int\dd{\tau}J^{++}_{2,ij}(\tau)\expval{T^{++}_{2,ij}(\tau)}_+=0
\end{equation}

\subsubsection{Two operators}

\paragraph{Cubic-Cubic:}

\begin{align}
    \noalign{\centering$\bigg\langle\displaystyle\int\dd{\tau_1}\dd{\tau_2}\displaystyle\frac{\lambda}{4N}\Tr{\tilde{\psi}_\alpha(\tau_1)(\gamma_i)_{\alpha\beta}\comm{\tilde{X}_i(\tau_1)}{\tilde{\psi}_\beta(\tau_1)}}\Tr{\tilde{\psi}_\gamma(\tau_2)(\gamma_j)_{\gamma\delta}\comm{\tilde{X}_j(\tau_2)}{\tilde{\psi}_\delta(\tau_2)}}\bigg\rangle_{+,c}=$}
    =-\frac{\lambda}{N}&\int\dd{\tau_1}\dd{\tau_2}A_\psi(\tau_1,\tau_2)\Tr{\comm{\tilde{X}^-_i(\tau_1)}{(\gamma_i)_{\alpha\beta}\tilde{\psi}^-_\beta(\tau_1)}\comm{\tilde{X}^-_j(\tau_2)}{(\gamma_j)_{\alpha\gamma}\tilde{\psi}^-_\gamma(\tau_2)}}+\nonumber\\
    +\frac{\lambda}{4N}&\int\dd{\tau_1}\dd{\tau_2}A_X(\tau_1,\tau_2)\Tr{(\gamma_i)_{\alpha\beta}\acomm{\tilde{\psi}^-_\alpha(\tau_1)}{\tilde{\psi}^-_\beta(\tau_2)}(\gamma_i)_{\gamma\delta}\acomm{\tilde{\psi}^-_\gamma(\tau_1)}{\tilde{\psi}^-_\delta(\tau_2)}}
\end{align}

\paragraph{Cubic-Quartic:}

\begin{gather}
    \bigg\langle-\int\dd{\tau_1}\dd{\tau_2}\qty(\frac{\lambda}{4N})^{3/2}\Tr{\tilde{\psi}_\alpha(\tau_1)(\gamma_i)_{\alpha\beta}\comm{\tilde{X}_i(\tau_1)}{\tilde{\psi}_\beta(\tau_1)}}\Tr{\comm{\tilde{X_j}(\tau_2)}{\tilde{X_k}(\tau_2)}^2}\bigg\rangle_{+,c}=\nonumber\\
    =\frac{1}{2}\qty(\frac{\lambda}{N})^{3/2}\int\dd{\tau_1}\dd{\tau_2}(\gamma_i)_{\alpha\beta}A_X(\tau_1,\tau_2)\Tr{\acomm{\tilde{\psi}^-_\alpha(\tau_1)}{\tilde{\psi}^-_\beta(\tau_1)}\comm{\tilde{X}^-_k(\tau_2)}{\comm{\tilde{X}^-_i(\tau_2)}{\tilde{X}^-_k(\tau_2)}}}
\end{gather}

\paragraph{Cubic-Source:}

\begin{align}
    \noalign{\centering$\bigg\langle-\displaystyle\int\dd{\tau_1}\dd{\tau_2}\displaystyle\sqrt{\frac{\lambda}{4N}}\Tr{\tilde{\psi}_\alpha(\tau_1)(\gamma_i)_{\alpha\beta}\comm{\tilde{X}_i(\tau_1)}{\tilde{\psi}_\beta(\tau_1)}}N^2J^{++}_{2,jk}(\tau_2)T^{++}_{2,jk}(\tau_2)\bigg\rangle_{+,c}=$}
    =-\sqrt{\frac{\lambda}{N}}\frac{1}{\lambda^{2/7}}\int\dd{\tau_1}\dd{\tau_2}A_X(\tau_1,\tau_2)\bigg[(\gamma_i)_{\alpha\beta}J^{++}_{2,(ij)}(\tau_2)&\Tr{\acomm{\tilde{\psi}^-_\alpha(\tau_1)}{\tilde{\psi}^-_\beta(\tau_1)}\tilde{X}^-_j(\tau_2)}-\nonumber\\
    -\frac{1}{9}(\gamma_i)_{\alpha\beta}J^{++}_{2,jj}(\tau_2)&\Tr{\acomm{\tilde{\psi}^-_\alpha(\tau_1)}{\tilde{\psi}^-_\beta(\tau_1)}\tilde{X}^-_i(\tau_2}\bigg]
\end{align}

\paragraph{Quartic-Quartic:}

\begin{align}
    \noalign{\centering$\expval{\displaystyle\frac{\lambda^2}{16N^2}\displaystyle\int\dd{\tau_1}\dd{\tau_2}\Tr{\comm{\tilde{X}_i(\tau_1)}{\tilde{X}_j(\tau_1)}^2}\Tr{\comm{\tilde{X}_k(\tau_2)}{\tilde{X}_l(\tau_2)}^2}}_{+,c}=$}
    =\frac{\lambda^2}{N^2}\int\dd{\tau_1}\dd{\tau_2}A_X(\tau_1,\tau_2)\Tr\bigg\{&\comm{\tilde{X}^-_i(\tau_1)}{\comm{\tilde{X}^-_j(\tau_1)}{\tilde{X}^-_i(\tau_1)}}\cdot\nonumber\\
    \cdot&\comm{\tilde{X}^-_k(\tau_2)}{\comm{\tilde{X}^-_j(\tau_2)}{\tilde{X}^-_k(\tau_2)}}\bigg\}
\end{align}

\paragraph{Quartic-Source:}

\begin{align}
    \noalign{\centering$\expval{-\displaystyle\frac{\lambda}{4N}\displaystyle\int\dd{\tau_1}\dd{\tau_2}\Tr{\comm{\tilde{X}_i(\tau_1)}{\tilde{X}_j(\tau_1)}^2}N^2J^{++}_{2,jk}(\tau_2)T^{++}_{2,jk}(\tau_2)}_{+,c}=$}
    =-\frac{\lambda^{5/7}}{N}\int\dd{\tau_1}\dd{\tau_2}A_X(\tau_1,\tau_2)J^{++}_{2,kl}(\tau_2)\Tr\bigg\{&\comm{\tilde{X}^-_i(\tau_1)}{\tilde{X}^-_k(\tau_1)}\comm{\tilde{X}^-_i(\tau_1)}{\tilde{X}^-_l(\tau_2)}-\nonumber\\
    -\frac{\delta_{kl}}{9}&\comm{\tilde{X}^-_i(\tau_1)}{\tilde{X}^-_j(\tau_1)}\comm{\tilde{X}^-_i(\tau_1)}{\tilde{X}^-_j(\tau_2)}\bigg\}
\end{align}

\paragraph{Source-Source:}

\begin{gather}
    \expval{\int\dd{\tau_1}\dd{\tau_2}N^2J^{++}_{2,ij}(\tau_2)T^{++}_{2,ij}(\tau_1)N^2J^{++}_{2,kl}(\tau_2)T^{++}_{2,kl}(\tau_2)}_{+,c}=\nonumber\\
    =\frac{4}{\lambda^{4/7}}\int\dd{\tau_1}\dd{\tau_2}A_X(\tau_1,\tau_2)J^{++}_{2,ij}(\tau_1)J^{++}_{2,kl}(\tau_2)\Tr\bigg\{\delta_{jk}\tilde{X}^-_i(\tau_1)\tilde{X}^-_l(\tau_2)-\nonumber\\
    -\frac{\delta_{kl}}{9}\tilde{X}^-_i(\tau_1)\tilde{X}^-_j(\tau_1)-\frac{\delta_{ij}}{9}\tilde{X}^-_k(\tau_1)\tilde{X}^-_l(\tau_2)+\frac{\delta_{ij}\delta_{kl}}{81}\tilde{X}^-_m(\tau_1)\tilde{X}^-_m(\tau_2)\bigg\}
\end{gather}

These results at first sight look quite daunting, as it does not look very clear how to interpret them. The main reason is that we have generated many new operators which were not there to begin with, and, some of them, violate supersymmetry (like the mass term). Something which we had already anticipated could happen due to the results from section \ref{subsec:smoothrg}. However, there is one simplifying aspect, there are no double trace operators.

Naively, this seems rather fatal. As we pointed out in section \ref{sec:overviewqrg}, if there are no double trace operators then there are no non-trivial dynamics for the new fields. This seems to be in stark contrast with the predictions in \cite{2000:Sekino} which predicts a non-trivial 2-point function for this mode, and with \cite{2011:Hanada} which checked it numerically. Note that it cannot be an artefact of us having neglected temperature since in \cite{2011:Hanada} finite temperature effects are also neglected and still they find non-trivial dynamics. It also cannot be an artefact over our choice of vacuum (namely, we expanded about the trivial vacuum) since in \cite{2011:Hanada} they used the same vacuum.

One could also worry that this is an artefact of the breaking of supersymmetry. However, our concern is that we have generated too few operators, if we had used a supersymmetry preserving regulator then the most that could have happened is a cancellation between separate diagrams, which would mean generating even fewer diagrams, which wouldn't solve this issue. Further, given that QRG only works if the regulator is spacetime dependent and \textit{all} local RG schemes break supersymmetry (as discussed in sections \ref{sec:rgbfss} and \ref{sec:lrg}), no scheme consistent with the original QRG proposal is capable of preserving supersymmetry, therefore this must be interpreted as a feature of the proposal itself.

However, there is still a possibility that we have missed something. The trick lies in the extra operators that we have generated\footnote{We thank Sung-Sik Lee for pointing this out}. If we don't have any reason to truncate them we should consider them in our analysis, however, the only way to do so seems to be adding sources for those operators in step 1. This by itself also goes against the supergravity predictions, this mode should have dynamics on its own, not just when coupled to other operators (and in the lattice simulations dynamics where observed without the need to turn on more operators). However, after we do this, we can take the limit where the original sources are all set to zero and then carry out the calculation anyway, possibly finding non-zero double trace operators which will only turn on away from the boundary. Then, technically, we have only turned on that single mode initially, it just so happened to turn on other modes which then gave it the necessary dynamics. This mechanism cannot be completely ruled out by our calculations, and it seems that our simplifying assumption that we only need to turn on a finite set of sources and still get meaningful answers is not justified, however, in practice, it is not possible (nor naively well defined) to turn on an infinite number of operators. This leads to many difficulties in proceeding and confirming or completely ruling out QRG, which the authors leave as open problems.

Firstly, it shouldn't be surprising that we have turned on extra modes, this is not a consistent truncation after all, this mode interacts with others. Therefore, in order to correctly interpret the results there should be some consistent way to truncate and neglect some operators to reproduce the approximation made in the supergravity side. However, neither the large $N$ limit nor generalised conformal dimensions seem to do the trick since all single trace operators scale equally in the large $N$ limit and in $d=1$ the fields have negative dimensions, so having more fields will lower the dimension even further.

To deal with this, one could try to use a consistent truncation instead. However, some of the single trace operators we have generated above are not part of the consistent truncation. This is problematic unless they never become dynamical. So we still run into the issue of having to turn on an infinite number of operators, with the added fact that we know that if QRG is valid then we can only generate double trace operators for those exact operators we turned on initially, we may still need an infinite number of auxiliary non-dynamical fields. The extent to which having those fields will affect physical results is unclear.

Finally, we note that, even in the case when no source is turned on, we still generate some single trace operators. None of these modes may at any point become dynamical because that would mean that the vacuum has non-trivial dynamics, which, once more, goes against the supergravity predictions. However, this is still not a full contradiction since it may be that these new modes are never dynamical unless we turn on sources at the start.\footnote{This also doesn't constrain the single-mode calculation too much, because even though the $\Tr(X^{2075})$ mode, for example, generated by the pure vacuum (modes of this form are eventually generated) cannot generate double-trace operators by itself or with other vacuum operators, it may still generate double trace operators when contracted with one of the modes turned on by the sources. So, even though by themselves they are non-dynamical we cannot just throw them away.}

This leads us with a very narrow window of possible success for QRG, it cannot generate any non-trivial dynamics when no source is turned on, it must generate non-trivial dynamics when any of the sources in \cite{2000:Sekino} is turned on, and it cannot generate non-trivial dynamics away from the consistent truncations in \cite{2013:Ortiz,2013:Anabalon,2014:Ortiz} when only those modes are initially turned on. Perhaps some clever use of $SO(9)$ symmetry could constrain which modes are turned on at each step and confirm or rule out QRG, however, currently the authors are unaware of any such method.

\section{Discussion}
There were three main steps in this paper: doing global RG on the BFSS model, developing a local RG scheme, and performing QRG on BFSS. The first two were part of the necessary construction to perform QRG, but they are also very important and interesting in their own right.

First of all, we performed standard Wilsonian RG on the BFSS model. This result was absent to the literature due to the finiteness of BFSS but was a very useful warm-up calculation. Even more importantly, it highlighted under which conditions were we able to preserve supersymmetry along the flow. Namely, a hard cutoff breaks supersymmetry but if we use Feynman parameters, as is usually done in higher dimensions, supersymmetry appears to be preserved. This is very surprising and the interpretation is not yet clear, because the physical hard cutoff breaks supersymmetry, the Feynman parametrisation is a mere computational trick. Furthermore, we concluded that the use of a smooth regulator always breaks supersymmetry. Even the use of the superspace formalism does not help because it does not preserve enough supersymmetry off-shell, it only preserves 4 supercharges out of the 16 total.

Secondly, we discussed under which conditions can we use a local regulator, and constructed an explicit example of one, a local Gaussian regulator. Constructing a local regulator is harder than a global one because of the subtleties of dealing with infinite dimensional objects, but we have shown that it is possible to do it, so long as we make sure every operator is symmetric and has an inverse. This section is especially interesting because it could potentially be used for performing RG in curved spacetime.

Finally, we put all the pieces together and performed QRG on BFSS with a particular operator turned on, one which we know from independent studies that has non-trivial dynamics in the gravity side, and found that it didn't generate any double trace operators. Further considerations meant it didn't completely rule out QRG but it greatly limited the ways in which it could still work. So far it appears to require turning on an infinite set of operators which is unclear if it is possible to do in practice. But further studies are necessary to fully understand its role in understanding the AdS/CFT correspondence.
\acknowledgments
We would like to thank Nick Dorey, Sung-Sik Lee, Enrico Pajer, and David Tong for reading an earlier version of this manuscript. J.~E.~S is supported in part by STFC grants PHY-1504541 and ST/P000681/1. J.~F.~Melo thanks the Cambridge Trust for his Vice-Chancellor's award to support his studies.
\appendix
\section{Details for the 1-loop calculation}\label{app:oneloop}

In this appendix we elaborate on the details of the 1-loop calculation from section \ref{subsec:hardrg}. We will often use identities for the structure constants of the algebra $\mathfrak{su}(N)$ taken from \cite{2017:Haber}.

\paragraph{Tadpole}

\begin{equation*}
\begin{tikzpicture}[baseline=(a)]
  \begin{feynman}
    \vertex (a) {$i,a$};
    \vertex [right=1.5cm of a] (centre);
    \vertex [right=0.5cm of centre] (aux1);
    \vertex [above=0.5cm of aux1] (aux2);
    \vertex [right=0.5cm of aux1] (aux3);
    \vertex [below=0.5cm of aux1] (aux4);
    
    \diagram* {
      (a) -- [scalar,momentum={[arrow shorten=0.27]$p$}] (centre),
      (centre) -- [blue,plain,out=90,in=180] (aux2),
      (aux2) -- [blue,plain,out=0,in=90,momentum={[arrow style=blue]$\omega$}] (aux3),
      (aux3) -- [blue,plain,out=-90,in=0] (aux4),
      (aux4) -- [blue,plain,out=180,in=-90] (centre),
    };
  \end{feynman}
  \draw[fill=black] (centre) circle (1.5pt);
\end{tikzpicture}
=\frac{\ii}{2}\sqrt{\frac{\lambda}{N}}(\gamma_i)_{\alpha\beta}f_{acb}\delta_{\alpha\beta}\delta_{bc}\int\frac{\dd{\omega}}{2\pi}\frac{1}{\omega}=0
\end{equation*}
given that the gamma matrices are traceless, $f_{abc}$ is totally anti-symmetric, and also because the integrand is odd (each of these statements individually would make this diagram vanish).

\paragraph{Scalar propagator}

\begin{equation*}
\begin{tikzpicture}[baseline=(b.south)]
  \begin{feynman}
    \vertex (a) {$i,a$};
    \vertex [right=1.5cm of a] (centre);
    \vertex [right=1.25cm of centre] (b) {$j,b$};
    \vertex [above=1.5cm of centre] (aux);
    
    \diagram* {
      (a) -- [scalar,out=0,in=180,momentum'={[arrow shorten=0.25] $p$}] (centre),
      (centre) -- [scalar,out=0,in=180,momentum'={[arrow shorten=0.25] $p$}] (b),
      (centre) -- [blue,scalar,out=135,in=180,momentum={[arrow style=blue, arrow shorten=0.27] $\omega$}] (aux),
      (aux) -- [blue,scalar,out=0,in=45] (centre),
    };
    \draw[fill=black] (centre) circle (1.5pt);
  \end{feynman}
\end{tikzpicture}
    \begin{array}{rl}
    =-\displaystyle\frac{\lambda}{2N}\Big[ & f_{abe}f_{cde}(\delta_{ik}\delta_{jl}-\delta_{il}\delta_{jk})+\\
    + & f_{ace}f_{bde}(\delta_{ij}\delta_{kl}-\delta_{il}\delta_{jk})+\\
    + & f_{ade}f_{bce}(\delta_{ij}\delta_{kl}-\delta_{ik}\delta_{jl})\Big]\delta_{kl}\delta_{cd}\displaystyle\int\frac{\dd{\omega}}{2\pi}\frac{1}{\omega^2}=
    \end{array}
\end{equation*}
\begin{align*}
    &=-\frac{\lambda}{2N}\qty[2N\delta_{ab}(9\delta_{ij}-\delta_{ij})+2N\delta_{ab}(9\delta_{ij}-\delta_{ij})]\int\frac{\dd{\omega}}{2\pi}\frac{1}{\omega^2}=\\
    &=-16\lambda\delta_{ab}\delta_{ij}\int\frac{\dd{\omega}}{2\pi}\frac{1}{\omega^2}
\end{align*}
using the identity $f_{abc}f_{abd}=2N\delta_{ab}$.

\begin{equation*}
\begin{tikzpicture}[baseline=(a)]
  \begin{feynman}
    \vertex (a) {$i,a$};
    \vertex [right=1.4cm of a] (centre1);
    \vertex [right=0.6cm of centre1] (aux1);
    \vertex [above=0.6cm of aux1] (aux2);
    \vertex [below=0.6cm of aux1] (aux3);
    \vertex [right=0.6cm of aux1] (centre2);
    \vertex [right=1.1cm of centre2] (b) {$j,b$};
    
    \diagram* {
      (a) -- [scalar,momentum'={[arrow shorten=0.25]$p$}] (centre1),
      (centre1) -- [blue,plain,out=90,in=180,momentum={[arrow style=blue,arrow shorten=0.27]$\omega$}] (aux2),
      (aux2) -- [blue,plain,out=0,in=90] (centre2),
      (centre2) -- [blue,plain,out=-90,in=0,momentum={[arrow style=blue,arrow shorten=0.27]$\omega-p$}] (aux3),
      (aux3) -- [blue,plain,out=180,in=-90] (centre1),
      (centre2) -- [scalar,momentum={[arrow shorten=0.25]$p$}] (b),
    };
  \end{feynman}
  \draw[fill=black] (centre1) circle (1.5pt);
  \draw[fill=black] (centre2) circle (1.5pt);
\end{tikzpicture}
=\frac{\lambda}{2N}(\gamma_i)_{\alpha\gamma}f_{aec}(\gamma_j)_{\beta\delta}f_{bfd}\delta_{cd}\delta_{\alpha\beta}\delta_{ef}\delta_{\gamma\delta}\int\frac{\dd{\omega}}{2\pi}\frac{1}{\omega(\omega-p)}=
\end{equation*}
\vspace{-10mm}
\begin{align}
    \hspace{1mm}=16\lambda\delta_{ab}\delta_{ij}\int\frac{\dd{\omega}}{2\pi}\frac{1}{\omega(\omega-p)}
\end{align}
using $f_{abc}f_{abd}=2N\delta_{ab}$ and $\tr(\gamma_i\gamma_j)=16\delta_{ij}$.

\paragraph{Fermion propagator}

\begin{equation*}
\begin{tikzpicture}[baseline=(a)]
  \begin{feynman}
    \vertex (a) {$\alpha,a$};
    \vertex [right=1.5cm of a] (centre1);
    \vertex [right=1.5cm of centre1] (centre2);
    \vertex [right=1.1cm of centre2] (b) {$\beta,b$};
    \vertex [right=0.75cm of centre1] (aux1);
    \vertex [above=0.75cm of aux1] (aux2);
    
    \diagram* {
      (a) -- [plain,momentum'={[arrow shorten=0.25]$p$}] (centre1),
      (centre1) -- [blue,plain,momentum'={[arrow style=blue,arrow shorten=0.25]$p-\omega$}] (centre2),
      (centre2) -- [plain,momentum'={[arrow shorten=0.25]$p$}] (b),
      (centre1) -- [blue,scalar,out=90,in=180,momentum={[arrow style=blue,arrow shorten=0.25]$\omega$}] (aux2),
      (aux2) -- [blue,scalar,out=0,in=90] (centre2),
    };
  \end{feynman}
  \draw[fill=black] (centre1) circle (1.5pt);
  \draw[fill=black] (centre2) circle (1.5pt);
\end{tikzpicture}
=-\frac{\lambda}{N}(\gamma_i)_{\alpha\gamma}f_{cea}(\gamma_j)_{\delta\beta}f_{dbf}\delta_{ij}\delta_{cd}\delta_{\gamma\delta}\delta_{ef}\int\frac{\dd{\omega}}{2\pi}\frac{1}{\omega^2(p-\omega)}=
\end{equation*}
\vspace{-5mm}
\begin{align}
    \hspace{10mm}=18\lambda\delta_{\alpha\beta}\delta_{ab}\int\frac{\dd{\omega}}{2\pi}\frac{1}{\omega^2(p-\omega)}
\end{align}
using $f_{abc}f_{abd}=2N\delta_{ab}$ and $\gamma_i\gamma_i=9$.

\paragraph{Triangle diagram}

\begin{equation*}
\begin{tikzpicture}[baseline=(a)]
  \begin{feynman}
    \vertex (a) {$i,a$};
    \vertex [right=1.5cm of a] (centre1);
    \vertex [right=1cm of centre1] (aux1);
    \vertex [above=0.75cm of aux1] (centre2);
    \vertex [below=0.75cm of aux1] (centre3);
    \vertex [right=1.25cm of centre2] (b) {$j,b$};
    \vertex [right=1.25cm of centre3] (c) {$k,c$};
    
    \diagram* {
      (a) -- [scalar,momentum={[arrow shorten=0.27]$p_1$}] (centre1),
      (centre1) -- [blue,plain,momentum={[arrow style=blue,arrow shorten=0.27]$\omega$}] (centre2),
      (centre2) -- [blue,plain,momentum={[arrow style=blue,arrow shorten=0.27]$\omega+p_2$}] (centre3),
      (centre1) -- [blue,plain,rmomentum'={[arrow style=blue,arrow shorten=0.27]$\omega-p_1$}] (centre3),
      (centre2) -- [scalar,rmomentum={[arrow shorten=0.27]$p_2$}] (b),
      (centre3) -- [scalar,rmomentum'={[arrow shorten=0.27]$p_3$}] (c),
    };
  \end{feynman}
  \draw[fill=black] (centre1) circle (1.5pt);
  \draw[fill=black] (centre2) circle (1.5pt);
  \draw[fill=black] (centre3) circle (1.5pt);
\end{tikzpicture}
=-\ii\qty(\frac{\lambda}{N})^{\frac{3}{2}}(\gamma_1)_{\alpha\beta}f_{aa_1d}(\gamma_j)_{\gamma\delta}f_{bfe}(\gamma_k)_{\epsilon\zeta}f_{chg}\int\frac{\dd{\omega}}{2\pi}\frac{\delta_{\alpha\gamma}\delta_{de}\delta_{\delta\epsilon}\delta_{fg}\delta_{\beta\zeta}\delta_{a_1h}}{\omega(\omega+p_2)(\omega-p_1)}=
\end{equation*}
\vspace{-15mm}
\begin{align*}
    \hspace{51mm}=-\ii\qty(\frac{\lambda}{N})^{\frac{3}{2}}\tr(\gamma_i\gamma_j\gamma_k)f_{ahd}f_{bfd}f_{chf}\int\frac{\dd{\omega}}{2\pi}\frac{1}{\omega(\omega+p_2)(\omega-p_1)}=0
\end{align*}
using $\tr(\gamma_i\gamma_j\gamma_k)=0$.

\paragraph{Cubic coupling}

\begin{equation*}
\begin{tikzpicture}[baseline=(c)]
  \begin{feynman}
    \vertex (c) {$i,c$};
    \vertex [right=1.75cm of c] (centre1);
    \vertex [right=0.9cm of centre1] (aux1);
    \vertex [right=1.1cm of aux1] (aux2);
    \vertex [above=0.65cm of aux1] (centre2);
    \vertex [below=0.675cm of aux1] (centre3);
    \vertex [above=1.2cm of aux2] (a) {$\alpha,a$};
    \vertex [below=1.2cm of aux2] (b) {$\beta,b$};

    \diagram* {
      (c) -- [scalar,momentum={[arrow shorten=0.25]$p_3$}] (centre1);
      (centre1) -- [blue,plain,rmomentum={[arrow style=blue,arrow shorten=0.25]$\omega$}] (centre2);
      (centre2) -- [plain,rmomentum={[arrow shorten=0.25]$p_1$}] (a);
      (centre1) -- [blue,plain,momentum'={[arrow style=blue,arrow shorten=0.25]$\omega+p_3$}] (centre3);
      (centre3) -- [plain,rmomentum'={[arrow shorten=0.25]$p_2$}] (b);
      (centre2) -- [blue,scalar,rmomentum={[arrow style=blue,arrow shorten=0.25]$\omega-p_1$}] (centre3);
    };
  \end{feynman}
  \draw[fill=black] (centre1) circle (1.5pt);
  \draw[fill=black] (centre2) circle (1.5pt);
  \draw[fill=black] (centre3) circle (1.5pt);
\end{tikzpicture}
=\ii\qty(\frac{\lambda}{N})^{\frac{3}{2}}(\gamma_i)_{\gamma\epsilon}f_{cgf}(\gamma_j)_{\zeta\beta}f_{a_1bh}(\gamma_k)_{\alpha\delta}f_{b_1da}\int\frac{\dd{\omega}}{2\pi}\frac{\delta_{\delta\gamma}\delta_{df}\delta_{\epsilon\zeta}\delta_{gh}\delta_{jk}\delta_{b_1a_1}}{\omega(\omega+p_3)(\omega-p_1)^2}=
\end{equation*}
\vspace{-20mm}
\begin{align}
    \hspace{48mm}=&\ii\qty(\frac{\lambda}{N})^{\frac{3}{2}}(\gamma_j\gamma_i\gamma_j)_{\alpha\beta}f_{aa_1d}f_{bga_1}f_{cgd}\int\frac{\dd{\omega}}{2\pi}\frac{1}{\omega(\omega+p_3)(\omega-p_1)^2}=\nonumber\\
    =&7\ii\lambda\sqrt{\frac{\lambda}{N}}(\gamma_i)_{\alpha\beta}f_{acb}\int\frac{\dd{\omega}}{2\pi}\frac{1}{\omega(\omega+p_3)(\omega-p_1)^2}
\end{align}
using $\gamma_j\gamma_i\gamma_j=-7\gamma_i$, and $f_{aa_1d}f_{bga_1}f_{cgd}=-Nf_{acb}$.

\paragraph{Quartic coupling}

\begin{equation*}
\begin{tikzpicture}[baseline=(d.south)]
  \begin{feynman}
    \vertex (a) {$i,a$};
    \vertex [below=2.25cm of a] (c) {$k,c$};
    \vertex [below=1.125cm of a] (aux1);
    \vertex [right=3.75cm of a] (b) {$j,b$};
    \vertex [right=3.75cm of c] (d) {$l,d$};
    \vertex [right=1.125cm of aux1] (centre1);
    \vertex [right=1.5cm of centre1] (centre2);
    
    \diagram* {
      (a) -- [scalar,out=-45,in=135,momentum={[arrow shorten=0.27]$p_1$}] (centre1),
      (c) -- [scalar,out=45,in=-135,momentum'={[arrow shorten=0.27]$p_3$}] (centre1),
      (centre1) -- [blue,scalar,out=90,in=90,looseness=1.7,momentum={[arrow style=blue,arrow shorten=0.4]$\omega$}] (centre2),
      (centre1) -- [blue,scalar,out=-90,in=-90,looseness=1.8,rmomentum'={[arrow style=blue,arrow shorten=0.4]$\omega+p_2+p_4$}] (centre2),
      (centre2) -- [scalar,out=45,in=-135,rmomentum={[arrow shorten=0.27]$p_2$}] (b),
      (centre2) -- [scalar,out=-45,in=135,rmomentum'={[arrow shorten=0.27]$p_3$}] (d),
    };
  \end{feynman}
  \draw[fill=black] (centre1) circle (1.5pt);
  \draw[fill=black] (centre2) circle (1.5pt);
\end{tikzpicture}
    \begin{array}{rl}
    =\displaystyle\frac{1}{2}\frac{\lambda^2}{N^2}\Big[ & f_{aea_1}f_{cga_1}(\delta_{ik}\delta_{mo}-\delta_{io}\delta_{mk})+\\
    + & f_{aca_1}f_{ega_1}(\delta_{im}\delta_{ko}-\delta_{io}\delta_{mk})+\\
    + & f_{aga_1}f_{eca_1}(\delta_{im}\delta_{ko}-\delta_{ik}\delta_{mo})\Big]\cdot\\
    \cdot\Big[ & f_{fbb_1}f_{hdb_1}(\delta_{np}\delta_{jl}-\delta_{nl}\delta_{jp})+\\
    + & f_{fhb_1}f_{bdb_1}(\delta_{nj}\delta_{pl}-\delta_{nl}\delta_{jp})+\\
    + & f_{fdb_1}f_{bhb_1}(\delta_{nj}\delta_{pl}-\delta_{np}\delta_{jl})\Big]\displaystyle\int\frac{\dd{\omega}}{2\pi}\frac{\delta_{mn}\delta_{ef}\delta_{op}\delta_{gh}}{\omega^2(\omega+p_2+p_4)^2}
    \end{array}
\end{equation*}

There is a bit of algebra in expanding all these terms and collecting them together, it involves using the fact that $f_{abc}$ is totally antisymmetric, $f_{aa_1d}f_{bga_1}f_{cgd}=-Nf_{acb}$, and some relabeling of indices. The end result for this diagram is,

\begin{align*}
    \frac{\lambda^2}{N^2}\Big[ & f_{ahe}f_{cef}f_{dfg}f_{bgh}(7\delta_{ik}\delta_{jl}+\delta_{ij}\delta_{kl})+\\
    + & f_{ahe}f_{cef}f_{bfg}f_{dgh}(7\delta_{ik}\delta_{jl}+\delta_{il}\delta_{jk})+\\
    + & 4Nf_{ace}f_{bde}(\delta_{ij}\delta_{kl}-\delta_{il}\delta_{jk})\Big] \int\frac{\dd{\omega}}{2\pi}\frac{1}{\omega^2(\omega+p_2+p_4)^2}
\end{align*}

Similarly, by making the substitutions $a\to b,~b\to d,~c\to a,~d\to c,~i\to j,~k\to l,~k\to i,~l\to k$ on the previous diagram, we get,

\begin{equation*}
\begin{tikzpicture}[baseline=(centre2)]
  \begin{feynman}
    \vertex (a) {$i,a$};
    \vertex [below=3.75cm of a] (c) {$k,c$};
    \vertex [right=2.25cm of a] (b) {$j,b$};
    \vertex [right=2.25cm of c] (d) {$l,d$};
    \vertex [right=1.125cm of a] (aux);
    \vertex [below=1.125cm of aux] (centre1);
    \vertex [below=1.5cm of centre1] (centre2);
    
    \diagram* {
      (a) -- [scalar,out=-45,in=135,momentum'={[arrow shorten=0.27]$p_1$}] (centre1),
      (c) -- [scalar,out=45,in=-135,momentum={[arrow shorten=0.27]$p_3$}] (centre2),
      (centre1) -- [scalar,out=45,in=-135,rmomentum'={[arrow shorten=0.27]$p_2$}] (b),
      (centre2) -- [scalar,out=-45,in=135,rmomentum={[arrow shorten=0.27]$p_4$}] (d),
      (centre1) -- [blue,scalar,out=180,in=180,looseness=1.7,rmomentum'={[arrow shorten=0.4,arrow style=blue]$\omega+p_3+p_4$}] (centre2),
      (centre1) -- [blue,scalar,out=0,in=0,looseness=1.7,momentum={[arrow shorten=0.4, arrow style=blue]$\omega$}] (centre2),
    };
  \end{feynman}
  \draw[fill=black] (centre1) circle (1.5pt);
  \draw[fill=black] (centre2) circle (1.5pt);
\end{tikzpicture}
    \begin{array}{rl}
    =\displaystyle\frac{\lambda^2}{N^2}\Big[ & f_{bhe}f_{aef}f_{cfg}f_{dgh}(7\delta_{ji}\delta_{lk}-\delta_{jl}\delta_{ik})+\\
    + & f_{bhe}f_{aef}f_{dfg}f_{cgh}(7\delta_{ji}\delta_{lk}+\delta_{jk}\delta_{li})+\\
    + & 4Nf_{bae}f_{dce}(\delta_{jl}\delta_{ik}-\delta_{jk}\delta_{li})\Big]\displaystyle\int\frac{\dd{\omega}}{2\pi}\frac{1}{\omega^2(\omega+p_3+p_4)^2}
    \end{array}
\end{equation*}

Now, by making $b\leftrightarrow d,~j\leftrightarrow l$ on the previous diagram, we get,
 
\begin{equation*}
\begin{tikzpicture}[baseline=(centre2)]
  \begin{feynman}
    \vertex (a) {$i,a$};
    \vertex [below=3.75cm of a] (c) {$k,c$};
    \vertex [right=3cm of a] (b) {$j,b$};
    \vertex [right=3cm of c] (d) {$l,d$};
    \vertex [right=1.125cm of a] (aux1);
    \vertex [below=1.125cm of aux1] (centre1);
    \vertex [below=1.5cm of centre1] (centre2);
    \vertex [left=0.5cm of b] (aux2);
    \vertex [below=0.5cm of aux2] (aux3);
    \vertex [left=0.5cm of d] (aux4);
    \vertex [above=0.6cm of aux4] (aux5);
    
    \diagram* {
      (a) -- [scalar,out=-45,in=135,momentum'={[arrow shorten=0.27]$p_1$}] (centre1),
      (c) -- [scalar,out=45,in=-135,momentum={[arrow shorten=0.27]$p_3$}] (centre2),
      (centre1) -- [scalar,out=15,in=125] (aux5),
      (aux5) -- [scalar,out=-55,in=125,rmomentum'={[arrow shorten=-0.25]$p_4$}] (d)
      (centre2) -- [scalar,out=-15,in=-125] (aux3),
      (aux3) -- [scalar,out=55,in=-135,rmomentum={[arrow shorten=-0.4]$p_2$}] (b)
      (centre1) -- [blue,scalar,out=180,in=180,looseness=1.7,rmomentum'={[arrow shorten=0.4, arrow style=blue]$\omega+p_2+p_3$}] (centre2),
      (centre1) -- [blue,scalar,out=0,in=0,looseness=1.7,momentum'={[arrow shorten=0.3,arrow style=blue]$\omega$}] (centre2),
    };
  \end{feynman}
  \draw[fill=black] (centre1) circle (1.5pt);
  \draw[fill=black] (centre2) circle (1.5pt);
\end{tikzpicture}
    \begin{array}{rl}
    =\displaystyle\frac{\lambda^2}{N^2}\Big[ & f_{dhe}f_{aef}f_{cfg}f_{bgh}(7\delta_{li}\delta_{jk}-\delta_{lj}\delta_{ik})+\\
    + & f_{dhe}f_{aef}f_{bfg}f_{cgh}(7\delta_{li}\delta_{jk}+\delta_{lk}\delta_{ji})+\\
    + & 4Nf_{dae}f_{bce}(\delta_{lj}\delta_{ik}-\delta_{lk}\delta_{ji})\Big]\displaystyle\int\frac{\dd{\omega}}{2\pi}\frac{1}{\omega^2(\omega+p_2+p_3)^2}
    \end{array}
\end{equation*}

Taken as it is, due to their different dependence on the external momenta, these diagrams do not add up nicely. However, if we set the external momentum to zero we get,

\begin{align*}
    \frac{\lambda^2}{N}\int\frac{\dd{\omega}}{2\pi}\frac{1}{\omega^4}\bigg\{4\Big[&f_{abe}f_{cde}(\delta_{ik}\delta_{jl}-\delta_{il}\delta_{jk})+\\
    +&f_{ace}f_{bde}(\delta_{ij}\delta_{kl}-\delta_{il}\delta_{jk})+\\
    +&f_{ade}f_{bce}(\delta_{ij}\delta_{kl}-\delta_{ik}\delta_{jl})\Big]+\\
    +8\Big[&f_{ahe}f_{cef}f_{dfg}f_{bgh}(\delta_{ik}\delta_{jl}+\delta_{ij}\delta_{kl})+\\
    +&f_{ahe}f_{cef}f_{bfg}f_{dgh}(\delta_{ik}\delta_{jl}+\delta_{il}\delta_{jk})+\\
    +&f_{ahe}f_{def}f_{cfg}f_{bgh}(\delta_{il}\delta_{jk}+\delta_{ij}\delta_{kl})\Big]\bigg\}
\end{align*}

Using the identity $f_{aef}f_{bfg}f_{cgh}f_{dhe}=4\delta_{ad}\delta_{bc}+2(\delta_{ab}\delta_{cd}+\delta_{ac}\delta_{bd})+\frac{1}{2}N(d_{ade}d_{bce}+f_{ade}f_{bce})$, where $d_{abc}=\Tr[\acomm{T^a}{T^b}T^c]$, and with a lot of patience, one gets,

\begin{align}
    \frac{\lambda^2}{N}\int\frac{\dd{\omega}}{2\pi}\frac{1}{\omega^4}\bigg\{6\Big[&f_{abe}f_{cde}(\delta_{ik}\delta_{jl}-\delta_{il}\delta_{jk})+\nonumber\\
    +&f_{ace}f_{bde}(\delta_{ij}\delta_{kl}-\delta_{il}\delta_{jk})+\nonumber\\
    +&f_{ade}f_{bce}(\delta_{ij}\delta_{kl}-\delta_{ik}\delta_{jl})\Big]+\nonumber\\
    +8\Big[&\delta_{ab}\delta_{cd}(5\delta_{ik}\delta_{jl}+5\delta_{il}\delta_{jk}+6\delta_{ij}\delta_kl)+\nonumber\\
    +&\delta_{ac}\delta_{bd}(6\delta_{ik}\delta_{jl}+5\delta_{il}\delta_{jk}+5\delta_{ij}\delta_{kl})+\nonumber\\
    +&\delta_{ad}\delta_{bc}(5\delta_{ik}\delta_{jl}+6\delta_{il}\delta_{jk}+5\delta_{ij}\delta_{kl})\Big]+\nonumber\\
    +2\Big[&d_{abe}d_{cde}(\delta_{ik}\delta_{jl}+\delta_{il}\delta_{jk}+2\delta_{ij}\delta_{kl})+\nonumber\\
    +&d_{ace}d_{bde}(\delta_{ij}\delta_{kl}+\delta_{il}\delta_{jk}+2\delta_{ik}\delta_{jl})+\nonumber\\
    +&d_{ade}d_{bce}(\delta_{ij}\delta_{kl}+\delta_{ik}\delta_{jl}+2\delta_{il}\delta_{jk})\Big]\bigg\}
    \label{eq:scalarloop}
\end{align}

The box diagrams require very similar computations.

\begin{equation*}
\begin{tikzpicture}[baseline=(aux2)]
  \begin{feynman}
    \vertex (a) {$i,a$};
    \vertex [right=3.75cm of a] (b) {$j,b$};
    \vertex [below=3.75cm of a] (c) {$k,c$};
    \vertex [right=3.75cm of c] (d) {$l,d$};
    \vertex [right=1.875cm of a] (aux1);
    \vertex [below=1.875cm of aux1] (aux2);
    \vertex [above=0.75cm of aux2] (aux3);
    \vertex [left=0.75cm of aux3] (centre1);
    \vertex [right=0.75cm of aux3] (centre2);
    \vertex [below=0.75cm of aux2] (aux4);
    \vertex [left=0.75cm of aux4] (centre3);
    \vertex [right=0.75cm of aux4] (centre4);
    
    \diagram* {
        (a) -- [scalar,momentum'={[arrow shorten=0.27]$p_1$}] (centre1),
        (b) -- [scalar,momentum={[arrow shorten=0.27]$p_2$}] (centre2),
        (c) -- [scalar,momentum={[arrow shorten=0.27]$p_3$}] (centre3),
        (d) -- [scalar,momentum'={[arrow shorten=0.27]$p_4$}] (centre4),
        (centre1) -- [blue,plain,momentum={[arrow shorten=0.27,arrow style=blue]$\omega$}] (centre2),
        (centre1) -- [blue,plain,rmomentum'={[arrow shorten=0.27,arrow style=blue]$\omega-p_1$}] (centre3),
        (centre2) -- [blue,plain,momentum={[arrow shorten=0.27,arrow style=blue]$\omega+p_2$}] (centre4),
        (centre3) -- [blue,plain,rmomentum'={[arrow shorten=0.27,arrow style=blue]$\omega+p_2+p_4$}] (centre4),
    };
  \end{feynman}
  \draw[fill=black] (centre1) circle (1.5pt);
  \draw[fill=black] (centre2) circle (1.5pt);
  \draw[fill=black] (centre3) circle (1.5pt);
  \draw[fill=black] (centre4) circle (1.5pt);
\end{tikzpicture}
    \begin{array}{rl}
    =-\displaystyle\frac{\lambda^2}{N^2} & (\gamma_i)_{\theta\alpha}f_{aed_1}(\gamma_j)_{\beta\gamma}f_{bgf}(\gamma_l)_{\delta\epsilon}f_{da_1h}(\gamma_k)_{\zeta\eta}f_{cc_1b_1}\cdot\\
    &\cdot\displaystyle\int\frac{\dd{\omega}}{2\pi}\frac{\delta_{\alpha\beta}\delta_{\gamma\delta}\delta_{\epsilon\zeta}\delta_{\eta\theta}\delta_{ef}\delta_{gh}\delta_{a_1b_1}\delta_{c_1d_1}}{\omega(\omega+p_2)(\omega+p_2+p_4)(\omega-p_1)}
    \end{array}
\end{equation*}

Using the identity $\tr(\gamma_i\gamma_j\gamma_l\gamma_k)=16(\delta_{ij}\delta_{lk}-\delta_{il}\delta_{jk}+\delta_{ik}\delta_{jl})$, one gets

\begin{align*}
    -\frac{16\lambda^2}{N^2}f_{ahe}f_{bef}f_{dfg}f_{cgh}(\delta_{ij}\delta_{kl}-\delta_{il}\delta_{jk}+\delta_{ik}\delta_{jl})\int\frac{\dd{\omega}}{2\pi}\frac{1}{\omega(\omega+p_2)(\omega+p_2+p_4)(\omega-p_1)}
\end{align*}

Similarly, by making the substitutions $b\leftrightarrow d$, $j\leftrightarrow l$ we get

\begin{equation*}
\begin{tikzpicture}[baseline=(aux2)]
  \begin{feynman}
    \vertex (a) {$i,a$};
    \vertex [right=3.75cm of a] (b) {$j,b$};
    \vertex [below=3.75cm of a] (c) {$k,c$};
    \vertex [right=3.75cm of c] (d) {$l,d$};
    \vertex [right=1.875cm of a] (aux1);
    \vertex [below=1.875cm of aux1] (aux2);
    \vertex [above=0.75cm of aux2] (aux3);
    \vertex [left=0.75cm of aux3] (centre1);
    \vertex [right=0.75cm of aux3] (centre2);
    \vertex [below=0.75cm of aux2] (aux4);
    \vertex [left=0.75cm of aux4] (centre3);
    \vertex [right=0.75cm of aux4] (centre4);
    \vertex [right=0.5625cm of centre2] (aux5);
    \vertex [below=1.3125cm of aux5] (aux6);
    \vertex [right=0.5625cm of centre4] (aux7);
    \vertex [above=1.3125cm of aux7] (aux8);
    
    \diagram* {
        (a) -- [scalar,momentum'={[arrow shorten=0.27]$p_1$}] (centre1),
        (b) -- [scalar,momentum={[arrow shorten=0.27]$p_2$}] (aux8),
        (aux8) -- [scalar] (centre4),
        (c) -- [scalar,momentum={[arrow shorten=0.27]$p_3$}] (centre3),
        (d) -- [scalar,momentum'={[arrow shorten=0.27]$p_4$}] (aux6),
        (aux6) -- [scalar] (centre2),
        (centre1) -- [blue,plain,momentum={[arrow shorten=0.27,arrow style=blue]$\omega$}] (centre2),
        (centre1) -- [blue,plain,rmomentum'={[arrow shorten=0.27,arrow style=blue]$\omega-p_1$}] (centre3),
        (centre2) -- [blue,plain,momentum'={[arrow shorten=0.27,arrow style=blue,arrow distance=0.2cm]$\omega+p_4$}] (centre4),
        (centre3) -- [blue,plain,rmomentum'={[arrow shorten=0.27,arrow style=blue]$\omega+p_2+p_4$}] (centre4),
    };
  \end{feynman}
  \draw[fill=black] (centre1) circle (1.5pt);
  \draw[fill=black] (centre2) circle (1.5pt);
  \draw[fill=black] (centre3) circle (1.5pt);
  \draw[fill=black] (centre4) circle (1.5pt);
\end{tikzpicture}
    \begin{array}{rl}
    =-\displaystyle\frac{16\lambda^2}{N^2} & f_{ahe}f_{def}f_{bfg}f_{cgh}(\delta_{il}\delta_{kj}-\delta_{ij}\delta_{lk}+\delta_{ik}\delta_{lj})\cdot\\
    &\cdot\displaystyle\int\frac{\dd{\omega}}{2\pi}\frac{1}{\omega(\omega+p_4)(\omega+p_2+p_4)(\omega-p_1)}
    \end{array}
\end{equation*}

Now substituting $a\leftrightarrow b$, $i\leftrightarrow j$ in the first diagram we get

\begin{equation*}
\begin{tikzpicture}[baseline=(aux2)]
  \begin{feynman}
    \vertex (a) {$i,a$};
    \vertex [right=3.75cm of a] (b) {$j,b$};
    \vertex [below=3.75cm of a] (c) {$k,c$};
    \vertex [right=3.75cm of c] (d) {$l,d$};
    \vertex [right=1.875cm of a] (aux1);
    \vertex [below=1.875cm of aux1] (aux2);
    \vertex [above=0.75cm of aux2] (aux3);
    \vertex [left=0.75cm of aux3] (centre1);
    \vertex [right=0.75cm of aux3] (centre2);
    \vertex [below=0.75cm of aux2] (aux4);
    \vertex [left=0.75cm of aux4] (centre3);
    \vertex [right=0.75cm of aux4] (centre4);
    \vertex [above=0.5625cm of centre2] (aux5);
    \vertex [left=1.3125cm of aux5] (aux6);
    \vertex [above=0.5625cm of centre1] (aux7);
    \vertex [right=1.3125cm of aux7] (aux8);
    
    \diagram* {
        (a) -- [scalar,momentum={[arrow shorten=0.27]$p_1$}] (aux6),
        (aux6) -- [scalar] (centre2),
        (b) -- [scalar,momentum'={[arrow shorten=0.27]$p_2$}] (aux8),
        (aux8) -- [scalar] (centre1),
        (c) -- [scalar,momentum={[arrow shorten=0.27]$p_3$}] (centre3),
        (d) -- [scalar,momentum'={[arrow shorten=0.27]$p_4$}] (centre4),
        (centre1) -- [blue,plain,momentum'={[arrow shorten=0.27,arrow style=blue]$\omega$}] (centre2),
        (centre1) -- [blue,plain,rmomentum'={[arrow shorten=0.27,arrow style=blue]$\omega-p_2$}] (centre3),
        (centre2) -- [blue,plain,momentum={[arrow shorten=0.27,arrow style=blue]$\omega+p_1$}] (centre4),
        (centre3) -- [blue,plain,rmomentum'={[arrow shorten=0.27,arrow style=blue]$\omega+p_1+p_4$}] (centre4),
    };
  \end{feynman}https://tex.stackexchange.com/questions/409824/how-to-adjust-momentum-arrows-in-tikz-feynman
  \draw[fill=black] (centre1) circle (1.5pt);
  \draw[fill=black] (centre2) circle (1.5pt);
  \draw[fill=black] (centre3) circle (1.5pt);
  \draw[fill=black] (centre4) circle (1.5pt);
\end{tikzpicture}
    \begin{array}{rl}
    =-\displaystyle\frac{16\lambda^2}{N^2} & f_{bhe}f_{aef}f_{dfg}f_{cgh}(\delta_{ji}\delta_{kl}-\delta_{jl}\delta_{ik}+\delta_{jk}\delta_{il})\cdot\\
    &\cdot\displaystyle\int\frac{\dd{\omega}}{2\pi}\frac{1}{\omega(\omega+p_1)(\omega+p_1+p_4)(\omega-p_2)}
    \end{array}
\end{equation*}

Performing similar manipulations as were done for the scalar loop diagrams we get, for the sum of these three box diagrams, when all external momenta is set to zero,

\begin{align}
    -\frac{\lambda^2}{N}\int\frac{\dd{\omega}}{2\pi}\frac{1}{\omega^4}\bigg\{8\Big[&f_{abe}f_{cde}(\delta_{ik}\delta_{jl}-\delta_{il}\delta_{jk})+\nonumber\\
    +&f_{ace}f_{bde}(\delta_{ij}\delta_{kl}-\delta_{il}\delta_{jk})+\nonumber\\
    +&f_{ade}f_{bce}(\delta_{ij}\delta_{kl}-\delta_{ik}\delta_{jl})\Big]+\nonumber\\
    +32\Big[&\delta_{ab}\delta_{cd}(2\delta_{ij}\delta_{kl}+\delta_{ik}\delta_{jl}+\delta_{il}\delta_jk)+\nonumber\\
    +&\delta_{ac}\delta_{bd}(2\delta_{ik}\delta_{jl}+\delta_{ij}\delta_{kl}+\delta_{il}\delta_{jk})+\nonumber\\
    +&\delta_{ad}\delta_{bc}(2\delta_{il}\delta_{jk}+\delta_{ij}\delta_{kl}+\delta_{ik}\delta_{jl})\Big]+\nonumber\\
    +8\Big[&d_{abe}d_{cde}\delta_{ij}\delta_{kl}+d_{ace}d_{bde}\delta_{ik}\delta_{jl}+d_{ade}d_{bce}\delta_{il}\delta_{jk}\Big]\bigg\}
    \label{eq:fermionloop}
\end{align}

Using the identity $f_{abe}f_{cde}=\frac{4}{N}(\delta_{ac}\delta_{bd}-\delta_{ad}\delta_{bc})+d_{ace}d_{bde}-d_{bce}d_{ade}$ to combine (\ref{eq:scalarloop}) and (\ref{eq:fermionloop}) we at last obtain the final answer

\begin{align}
    -4\frac{\lambda^2}{N}\int\frac{\dd{\omega}}{2\pi}\frac{1}{\omega^4}\Big[&f_{abe}f_{cde}(\delta_{ik}\delta_{jl}-\delta_{il}\delta_{jk})+\nonumber\\
    +&f_{ace}f_{bde}(\delta_{ij}\delta_{kl}-\delta_{il}\delta_{jk})+\nonumber\\
    +&f_{ade}f_{bce}(\delta_{ij}\delta_{kl}-\delta_{ik}\delta_{jl})\Big]
\end{align}


\bibliographystyle{JHEP.bst}
\bibliography{bibfile}



\end{document}